\documentclass[lettersize,journal]{IEEEtran}
\usepackage{amsmath,amsfonts}
\usepackage{algorithmic}
\usepackage{algorithm}
\usepackage{array}
\usepackage[caption=false,font=normalsize,labelfont=sf,textfont=sf]{subfig}
\usepackage{textcomp}
\usepackage{stfloats}
\usepackage{url}
\usepackage{verbatim}
\usepackage{graphicx}
\usepackage{cite}
\usepackage{color,array,amsthm}
\usepackage{graphicx}
\usepackage{amsmath}
 \usepackage{amssymb}
\usepackage{float}

\newtheorem{theorem}{Theorem}

\begin{document}

\author{IEEE Publication Technology,~\IEEEmembership{Staff,~IEEE,}
\thanks{This paper was produced by the IEEE Publication Technology Group. They are in Piscataway, NJ.}
\thanks{Manuscript received April 19, 2021; revised August 16, 2021.}}

\markboth{Journal of \LaTeX\ Class Files,~Vol.~14, No.~8, August~2021}%
{Shell \MakeLowercase{\textit{et al.}}: A Sample Article Using IEEEtran.cls for IEEE Journals}

\title{Complex Aircraft Maneuvering using Reinforcement-Learning-Augmented Sliding Mode-Based Control}

\author{
Imran Sayyed,
Nandan Kumar Sinha%
\thanks{
Imran Sayyed is a M.S. Scholar with the Department of Aerospace Engineering,
Indian Institute of Technology Madras, Chennai 600036, India
(e-mail: imransayyed23@gmail.com).}
\thanks{
Nandan Kumar Sinha is a Professor with the Department of Aerospace Engineering,
Indian Institute of Technology Madras, Chennai 600036, India
(e-mail: nksinha@iitm.ac.in).}
}

\maketitle
\thispagestyle{empty}
\pagestyle{empty}
\begin{abstract}
Learning-based controllers leverage nonlinear couplings and enhance transients but seldom offer guarantees under tight input constraints. Robust feedback like sliding-mode control (SMC) provides these guarantees but is conservative in isolation. This paper creates a learning-augmented framework where a deep reinforcement learning policy produces learned nominal maneuvering commands and an SMC law imposes actuator limits, bounds learned authority, and guarantees robustness. The policy is modeled as a matched, bounded input, and Lyapunov-based conditions link SMC gains to the nominal compensation mismatch, guaranteeing stability under saturation. This formulation is applicable to nonlinear, underactuated plants with hard constraints. To illustrate the methodology, the method is applied to a six-degree-of-freedom aircraft model and compared with Reinforcement Learning and isolated SMC. Simulation results show that the hybrid controller improves transient behavior and reduces control oscillations compared to standalone RL and SMC controllers. A Monte Carlo evaluation over 1000 randomized initial flight conditions further confirms that the controller consistently converges to the target attitude with small terminal tracking error, demonstrating robustness to variation in the initial-condition envelope. Even using it with partially trained policies, SMC component of the control stabilizes transients, whereas fully trained policies provide faster convergence, reduced constraint violations, and robustness. These results illustrate that learning-augmented control offers superior performance with robustness guarantees under tight input constraints.
\end{abstract}


\section{INTRODUCTION}

Aircraft control across a broad flight envelope presents significant challenges due to the nonlinear nature of the underlying physics, high coupling between axes, and hard actuator constraints. As the vehicle transitions from low-angle, low-rate flight to high angles of attack or aggressive maneuvers, aerodynamic moments and forces become highly dependent on rate and attitude. In such conditions, control surface effectiveness varies with the state, and small control inputs can lead to large, coupled motions in roll, pitch, and yaw. Additionally, phenomena like aerodynamic hysteresis or flow separation make the input–output relationship far from linear. Practical implementations of controllers also face constraints such as actuator rate limits, time delays, and saturation, all of which must be considered. During recovery maneuvers, the throttle is often fixed or manually adjusted, limiting the number of rapidly changing inputs and capturing engine dynamics, which leaves fewer independent control channels relative to the number of states.

In the past decade and a half, hybrid control architectures that integrate reinforcement learning (RL) with classical or modern feedback controllers have gained considerable attention across aerospace, robotics, and general nonlinear systems. Early works combined actor–critic RL with robust feedback laws such as RISE to achieve asymptotic tracking with theoretical stability guarantees in nonlinear systems \cite{bhasin2011asymptotic}. Subsequent research explored RL augmentation of PID controllers, where the learning agent adjusts gains or generates nominal commands to improve tracking and robustness without compromising stability \cite{hao2025ddpg, sierra2022wind}. Similarly, RL has been combined with LQR to provide adaptive or residual control, leveraging LQR’s optimality and stability while allowing RL to compensate for model uncertainties or nonlinear effects \cite{baek2022hybrid,zanon2020safe}. Model predictive control (MPC) has also been integrated with RL, either as a safety filter to enforce constraints or as a parameterized optimizer tuned by RL, enabling performance improvement while ensuring constraint satisfaction \cite{zanon2020safe}. 

Sliding-mode control (SMC) has been paired with RL to combine robust finite-time convergence and disturbance rejection with learning-based adaptation for enhanced trajectory tracking \cite{dao2021adaptive}. Independently of learning-based augmentation, the SMC literature itself has continued to refine how the sliding surface and feedback gain are designed under uncertainty and actuator limits. On the sliding-surface side, recent work has progressed from finite-time convergent sliding functions to designs offering a designer-specified upper bound on convergence time, and to constraint-aware formulations for uncertain systems explicitly subject to input saturation, evaluated with a switching-form control law and Lyapunov-based ultimate-boundedness guarantees, and benchmarked against both an adaptive-robust controller and a conventional SMC baseline \cite{he2025taskspace,shao2025slidingreview}. On the gain-adaptation side, adaptive-gain sliding-mode designs have been proposed that estimate an unknown disturbance or mismatch bound online, rather than assuming it a priori, using either a leakage-type differential update law or an algebraic class-$\mathcal{K}$/$\mathcal{K}_\infty$-type law that predefines the sliding variable's convergence region independently of the true disturbance magnitude \cite{shao2025slidingreview}. It is also worth distinguishing the present architecture from other recent RL-SMC hybrids specifically. In several such hybrids, the learning component is used to tune a scalar parameter of an otherwise conventional SMC law, for example, adaptively shrinking the SMC boundary-layer width by estimating the uncertainty bound online \cite{youssef2025arsmc}, or learning the shape of the SMC reaching law to reduce chattering, validated on a real quadrotor \cite{wang2023chattering}, while the sliding-mode law itself continues to generate the control command. In contrast, the present architecture assigns the RL policy the role of generating the primary maneuvering command, with the SMC term acting as a bounded corrective feedback layer rather than the nominal command generator; this feedforward role, together with the explicit tunable authority factor $\zeta$ and the associated Lyapunov-based mismatch-bound analysis, is what is extended relative to our own earlier work in \cite{sayyed2026deepreinforcementlearningbased}. These developments are complementary to, rather than competing with, the present RL-augmented approach: they address how the SMC gain, sliding surface, and disturbance-bound estimate can themselves be refined under saturation and uncertainty, whereas this work addresses how a learned feedforward term is bounded and integrated alongside a fixed-structure SMC feedback law; the online, bound-free gain-adaptation techniques above are a natural direction for relaxing the a priori mismatch-bound assumption used in the stability analysis of Section~V. More recently, feedforward–feedback architectures have emerged, in which RL learns anticipatory nominal control actions that act as feedforward compensation while a conventional feedback controller ensures stability and enforces constraints \cite{vanhoof2023table}.

In this context, the goal of this paper is to design a controller for the post-stall spin recovery of a six-degree-of-freedom (6-DOF) F-18 model under conditions of input saturation, model uncertainty, and cross-axis coupling. While multiple control methods, such as sliding-mode control (SMC) and reinforcement learning (RL), can be used to address these challenges, each comes with its own set of trade-offs. This paper significantly extends our earlier work in \cite{sayyed2026deepreinforcementlearningbased} by introducing a bounded learning-based nominal compensation for learning-based control, a tunable authority factor linking learned and robust control actions, and a Lyapunov-based stability analysis under actuator saturation.

Robust controllers like SMC are known for their stability guarantees but tend to be conservative, especially in highly nonlinear regimes. On the other hand, data-driven controllers such as RL excel in handling complex dynamics but often lack formal safety guarantees and robustness in the presence of disturbances and uncertainties \cite{Sutton2018}. This motivates the integration of both methods into a hybrid controller that combines the strengths of an RL with the stability and robustness provided by SMC \cite{stough1985stall,croom1993spin}.

The key contributions of this work are as follows:
\begin{itemize}
    \item This work proposes a hybrid control methodology that combines a reinforcement learning (RL)-based learned nominal compensation component with a sliding mode controller (SMC), providing high-performance maneuver generation while ensuring robustness under hard actuator limits.
    
    \item A safety-filtered RL design is introduced, allowing the learning-based nominal compensation policy to operate within certified bounds while preserving the stability guarantees provided by the robust SMC feedback.
    \item The effectiveness of the hybrid controller is demonstrated in a challenging post-stall recovery scenario for a six-degree-of-freedom F-18 model, with comparisons drawn against standalone RL and SMC baselines.
    \item An implementation-ready framework is provided for integrating RL and SMC, enabling straightforward application to a wide range of platforms beyond the F-18 case study.
\end{itemize}

Although the RL policy is state-dependent, its role within the proposed architecture is that of a nominal control generator rather than a corrective feedback controller. The learned policy produces the primary maneuvering command by approximating the nominal inverse dynamics and coordinated nonlinear couplings of the aircraft, while the SMC compensates the residual mismatch, disturbances, and model uncertainties. Consequently, the RL pathway is treated as a learned feedforward compensation channel, whereas robustness and closed-loop stability are provided by the SMC feedback loop.

The remainder of this paper is organized as follows: Section II introduces the aircraft model and control task, defining the available state variables, control channels, and the constraints that must be considered during the control design. Section III discusses the RL framework used in the study, detailing the environment setup, observation and action spaces, reward shaping, and training configuration. Section IV presents the sliding-mode controller design, including methods for minimizing chattering and ensuring robustness against disturbances and model uncertainties. Section V explains how the hybrid RL-SMC controller is formulated, combining the RL component with the robust feedback from SMC under a simple coordination policy. Finally, Section VI presents results comparing the hybrid controller to RL-only and SMC-only controllers, demonstrating its performance improvements in terms of stability, convergence, and actuator usage.

\section{Model Description}

\begin{figure}
    \centering
    \includegraphics[width=1\linewidth]{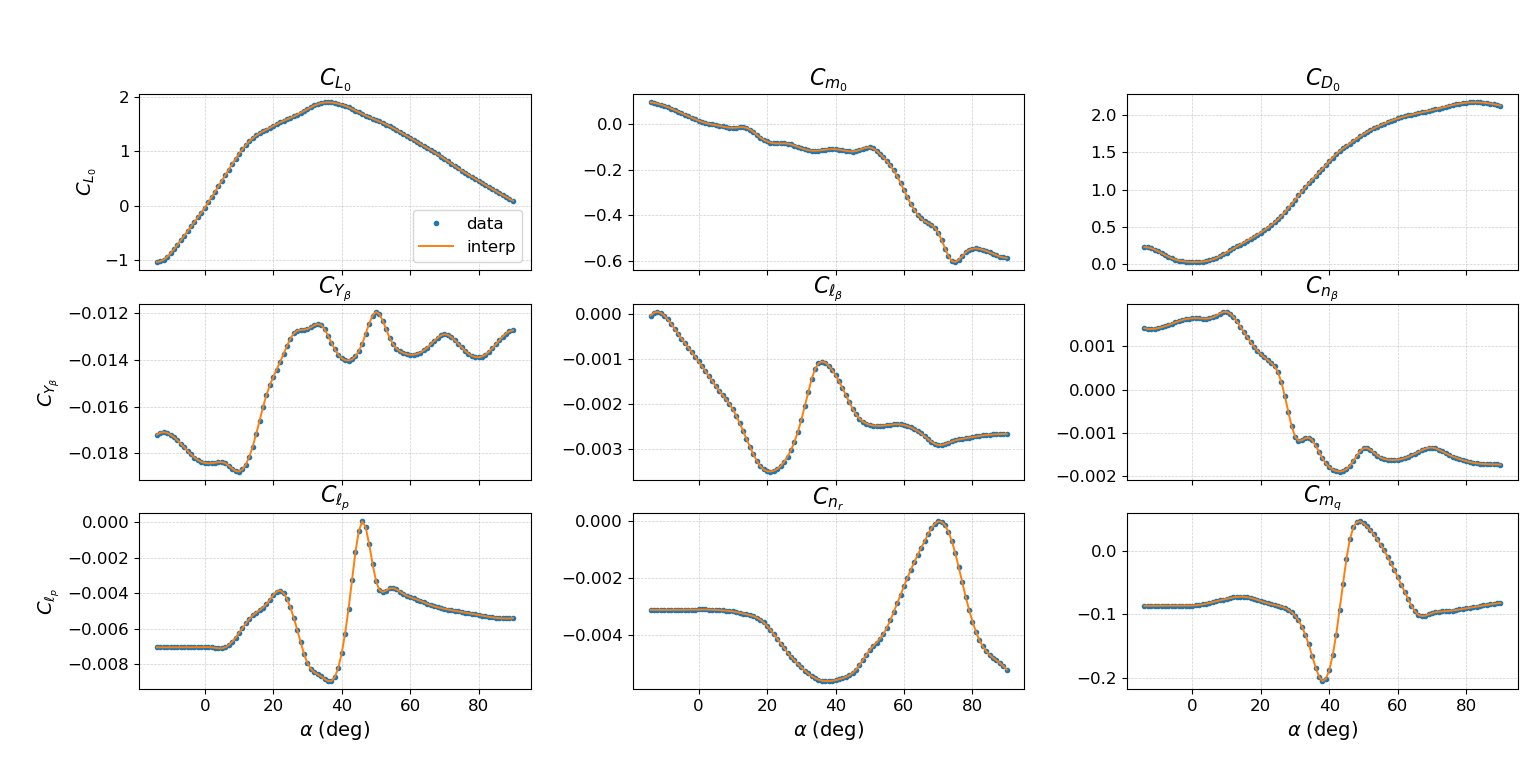}
    \caption{Key coefficients and stability derivatives vs $\alpha$.}
    \label{fig:st_dr_alpha}
\end{figure}

\begin{figure}
    \centering
    \includegraphics[width=1\linewidth]{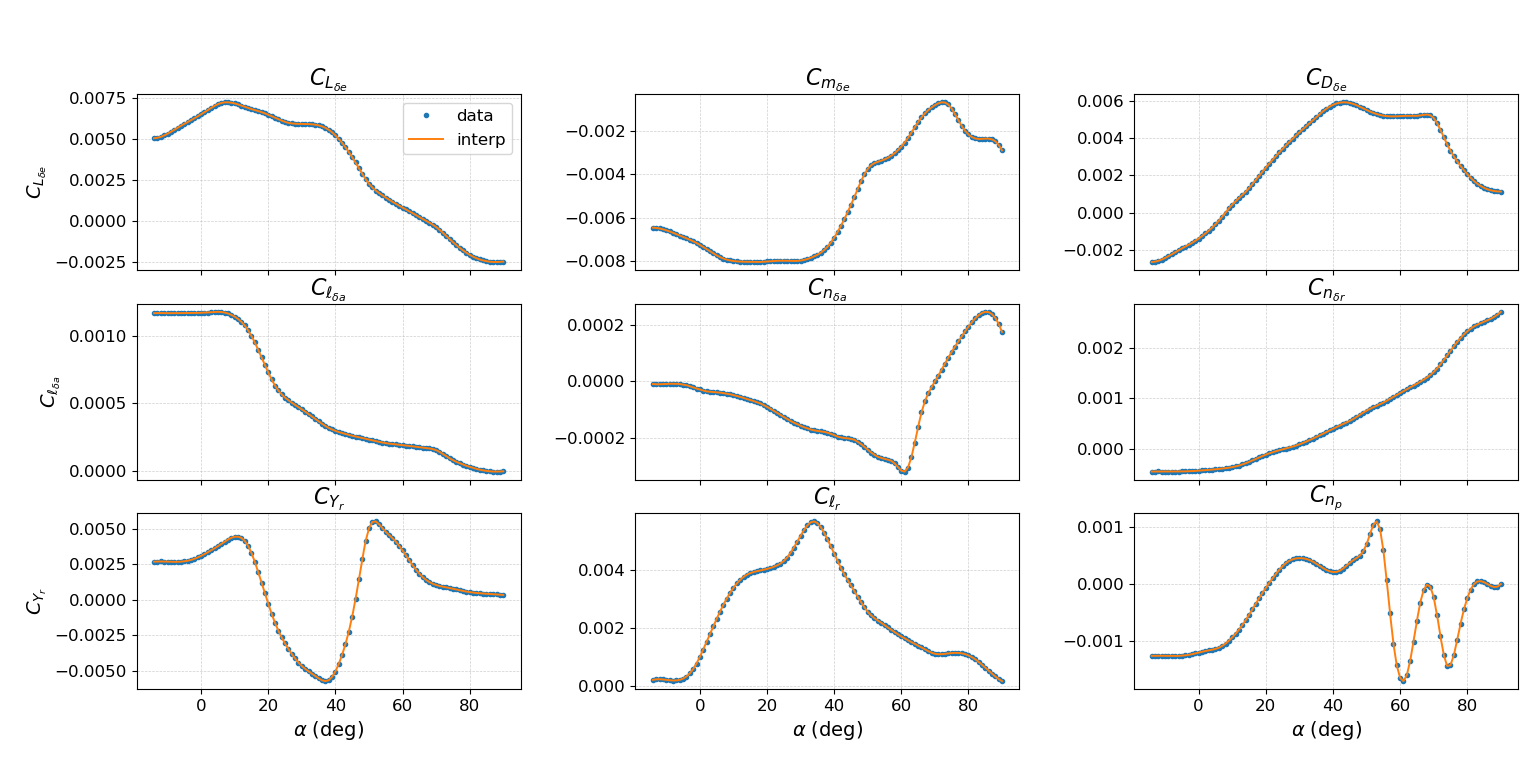}
    \caption{Control coupling and dynamic derivatives vs $\alpha$.}
    \label{fig:ct_dr_alpha}
\end{figure}

 This work considers a nonlinear model of the F-18/HARV, which exhibits many loss of control scenario via nonlinear phenomena. Aerodynamic data for this model is available for angle of attack range of -0.24 radians to 1.57 radians ($-15^o$ to $90^o$). Some key stability and control derivative data plots are shown in Fig. \ref{fig:st_dr_alpha} and Fig. \ref{fig:ct_dr_alpha}. Aerodynamic coefficients as function of state and control derivatives and state variables can be represented as:

\begin{small}
\begin{align*}
C_L &= C_{L_0} + C_{L_q}\,\frac{q\,\bar{c}}{2V} + C_{L_{\delta e}}\,\delta_e,\\[4pt]
C_D &= C_{D_0} + C_{D_q}\,\frac{q\,\bar{c}}{2V} + C_{D_{\delta e}}\,\delta_e,\\[4pt]
C_m &= C_{m_0} + C_{m_q}\,\frac{q\,\bar{c}}{2V} + C_{m_{\delta e}}\,\delta_e,\\[6pt]
C_Y &= C_{Y_\beta}\,\beta + C_{Y_p}\,\frac{p\,b}{2V} + C_{Y_r}\,\frac{r\,b}{2V}
      + C_{Y_{\delta e}}\,\delta_e + C_{Y_{\delta a}}\,\delta_a  \\ & \ \ \ + C_{Y_{\delta r}}\,\delta_r,\\[4pt]
C_l &= C_{l_\beta}\,\beta + C_{l_p}\,\frac{p\,b}{2V} + C_{l_r}\,\frac{r\,b}{2V}
      + C_{l_{\delta e}}\,\delta_e + C_{l_{\delta a}}\,\delta_a \\ & \ \ \ + C_{l_{\delta r}}\,\delta_r,\\[4pt]
C_n &= C_{n_\beta}\,\beta + C_{n_p}\,\frac{p\,b}{2V} + C_{n_r}\,\frac{r\,b}{2V}
      + C_{n_{\delta e}}\,\delta_e + C_{n_{\delta a}}\,\delta_a \\ & \ \ \ + C_{n_{\delta r}}\,\delta_r.
\end{align*}
\end{small}

The 6-DOF rigid body flight system governing equations based on \cite{r6} are given from \eqref{eq:vel} to \eqref{eq:chi}.

\begin{small}
\begin{align}
\dot{V} &= \frac{1}{m} \left[ T_m \eta \cos \alpha \cos \beta - \frac{1}{2} \rho V^2 S C_D(\alpha, q, \delta_e) - mg \sin \gamma \right] \label{eq:vel}\\
\dot{\alpha} &= q - \frac{1}{\cos \beta} \left[ (p \cos \alpha + r \sin \alpha) \sin \beta \right] \notag \\
& \quad + \frac{1}{m V} \left\{ T_m \eta \sin \alpha + \frac{1}{2} \rho V^2 S C_L(\alpha, q, \delta_e) - mg \cos \mu \cos \gamma \right\}  \label{eq:alpha} \\
\dot{\beta} &= \frac{1}{m V} \left[ - T_m \eta \cos \alpha \sin \beta + \frac{1}{2} \rho V^2 S C_Y(\alpha, \beta, p, r, \delta_e, \delta_a, \delta_r) \right] \notag \\
& \quad + mg \sin \mu \cos \gamma + (p \sin \alpha - r \cos \alpha)  \label{eq:beta} \\
\dot{p} &= \frac{I_y - I_z}{I_x} qr + \frac{1}{2 I_x} \rho V^2 S b C_l(\alpha, \beta, p, r, \delta_e, \delta_a, \delta_r) \label{eq:p}\\
\dot{q} &= \frac{I_z - I_x}{I_y} pr + \frac{1}{2 I_y} \rho V^2 S c C_m(\alpha, q, \delta_e) \label{eq:q}\\
\dot{r} &= \frac{I_x - I_y}{I_z} pq + \frac{1}{2 I_z} \rho V^2 S b C_n(\alpha, \beta, p, r, \delta_e, \delta_a, \delta_r) \label{eq:r}
\end{align}
\end{small}

\begin{small}
\begin{align}
\dot{\mu} &= \sec \beta (p \cos \alpha + r \sin \alpha)
+ \frac{1}{m V} \bigg\{
\frac{1}{2} \rho V^2 S C_L \tan \beta \\ \notag
&\quad + T_m \eta \sin \alpha
+ \sin \mu \tan \gamma
- mg \cos \mu \cos \gamma \tan \beta \\ \notag
&\quad + \frac{1}{2} \rho V^2 S C_Y \cos \mu \tan \gamma
\bigg\}
\label{eq:mu}
\end{align}
\end{small}

\begin{small}
\begin{align}
\dot{\gamma} &= \frac{1}{m V} \left\{ T_m \eta \left( \sin \alpha \cos \mu + \cos \alpha \sin \beta \sin \mu \right) \right. \\ \notag
& \quad \left. - \frac{1}{2} \rho V^2 S C_L \cos \mu - mg \cos \gamma - \frac{1}{2} \rho V^2 S C_Y \sin \mu \right\}   \label{eq:gamma} 
\end{align} 
\end{small}

\begin{small}
\begin{align}
\dot{\chi} &= \frac{1}{m V \cos \gamma} \left\{ T_m \eta \left( \sin \alpha \sin \mu - \cos \alpha \sin \beta \cos \mu \right) \right. \\ \notag 
& \quad \left. + \frac{1}{2} \rho V^2 S \left( C_L \sin \mu + C_Y \cos \mu \right) \right\} 
\label{eq:chi}
\end{align} 
\end{small}

The model can be represented in the nonlinear state-space form as:

\begin{align}
    \dot{x} = f(x) + g(x) sat(u)
\end{align}
  
Where $x$ are the states and $u$ are the control inputs of the aircraft model.
$$
x(t) = [V, \alpha, \beta, p, q, r, \mu, \gamma]
$$

$$
u(t) = [\eta, \delta _e, \delta _a, \delta _r ].
$$

\begin{figure*}[h]
    \centering
\includegraphics[width=\textwidth,height=\textheight,keepaspectratio]{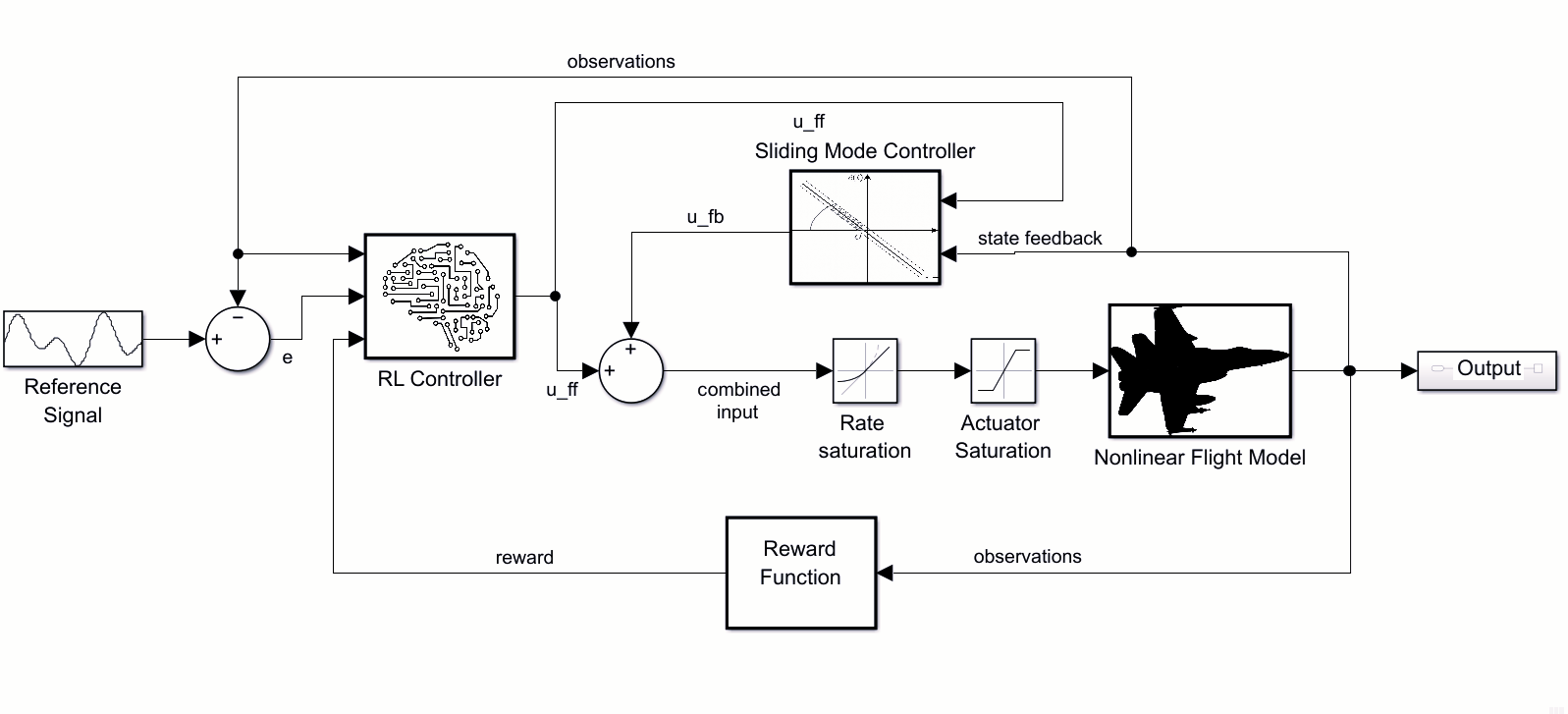} 
    \caption{ A schematic block diagram of proposed controller using integration of the RL and SMC control.}
    \label{fig:Block_Diagram3}
\end{figure*}

This is an underactuated system. Here, the $sat(u)$ function models the constraints on the input term. For the vector control input $u = [u_1, u_2, u_3]^\top \in \mathbb{R}^3$, the
saturation function is applied element-wise, i.e.,
$sat(u) = [sat(u_1), sat(u_2), sat(u_3)]^\top$, where for each scalar
component $u_i$,
\[
sat(u_i) =
\left\{
\begin{array}{ll}
  u_{i,max} & \text{if } u_i > u_{i,max} \\
  u_i & \text{if } u_{i,min} \leq u_i \leq u_{i,max} \\
  u_{i,min} & \text{if } u_i < u_{i,min}
\end{array}
\right.
\]
with channel-specific limits $u_{i,min}, u_{i,max}$ corresponding to the
elevator, aileron, and rudder deflection bounds given in Table
\ref{tab:obs_action_space}.

As the work presented in this paper is related to the faster dynamics, slower dynamics input $\eta$ is kept fixed at a constant value. Only control inputs used are $u \equiv [\delta _e, \delta _a, \delta _r],$  $u \in \mathbf{R}^3$.

\section{Control Methods}

This section illustrates three control elements that complement each other and how they are combined. The first section explains reinforcement learning, an incremental decision-making method. It learns an interaction-based policy mapping states to commands. Proximal Policy Optimization helps make gradual, conservative changes that respect the rules. The second section explains sliding mode control as a robust feedback method. It describes the sliding objective, applying input-output linearization to focus on controlled areas, and how to reduce unwanted oscillations by using boundary layers or higher-order designs without sacrificing actuator movement and speed limits. The third section describes a combination where the learned policy is used initially to support fast response and coordination between axes. Fig. \ref{fig:Block_Diagram3} shows the block diagram of the architecture of the proposed controller. The trained policy tries to keep the system at safe limits while the sliding-mode law closes the loop to ensure precision and reliability despite uncertainty. All of these sections in aggregate represent a design plan: stabilize rates before achieving accurate position recovery, specify constraints clearly, and train using varied scenarios. They also show how all this is combined into a ready-to-fly system \cite{Schulman2017PPO,Garcia2015SafeRLSurvey}.

\subsection{Reinforcement Learning}
Reinforcement learning (RL) deals with sequential decision-making problems where there are no labeled targets but only feedback in the form of reward-only. An agent continues to interact with an environment, from a given state it selects an action, the environment returns a new state and a reward, and learning is done by trial and error. Improvement is based on a trade-off between exploration (trying something else in order to get information) and exploitation (doing what appears to be working at the moment) so as to maximize long-term return.

When framed as a Markov Decision Process (MDP), good behavior is described in terms of Bellman optimality relations \cite{bellman1957dynamic}. The behavior of the agent is captured by a policy $\pi(a \mid \bar{s})$, i.e., a control law that specifies a probability for every action in a state. Learning operates to refine this policy to an optimal policy, $\pi^{*}(a\mid \bar{s})$, that maximizes cumulative expected reward. Value functions are used to estimate states or state–action pairs under a policy, and there are various families of algorithms that use them in different manners. Value-based methods estimate them and act greedily, policy-based methods update policy parameters directly and the actor–critic and other hybrids do both. These approaches have been found beneficial in robotics, games, networked systems, and autonomous platforms \cite{dao2021adaptive, vanhoof2023table}.

Proximal Policy Optimization (PPO) is an on-policy method which says the policy utilized to sample data is the one to be updated, eliminating the necessity of a replay buffer employed by off-policy methods such as DQN or DDPG. PPO follows a stochastic policy, samples new trajectories employing the current policy for each training iteration, and subsequently updates that policy based on the recently sampled on-policy batch. This is finalized to optimize the policy and maintain updates stable and conservative \cite{Lillicrap2016DDPG}.

Proximal Policy Optimization (PPO) promotes exploration by adding an entropy term to its loss while keeping updates small enough for stable learning. It was designed as a practical alternative to Trust Region Policy Optimization (TRPO): instead of solving a heavy constrained problem each step, PPO uses simple surrogate objectives that approximate a trust region. Two variants are common, one penalizes deviations with a KL term, and the other “clips” the probability ratio. The clipped form is widely used because it curbs overly large policy shifts and keeps training steady. Introduced by Schulman and colleagues, PPO delivers reliable, sample-efficient policy improvements, addressing the instability that can arise when classic policy-gradient methods take steps that are too large \cite{Schulman2015TRPO}.

\begin{figure}[H]
    \centering
    \includegraphics[width=1\linewidth]{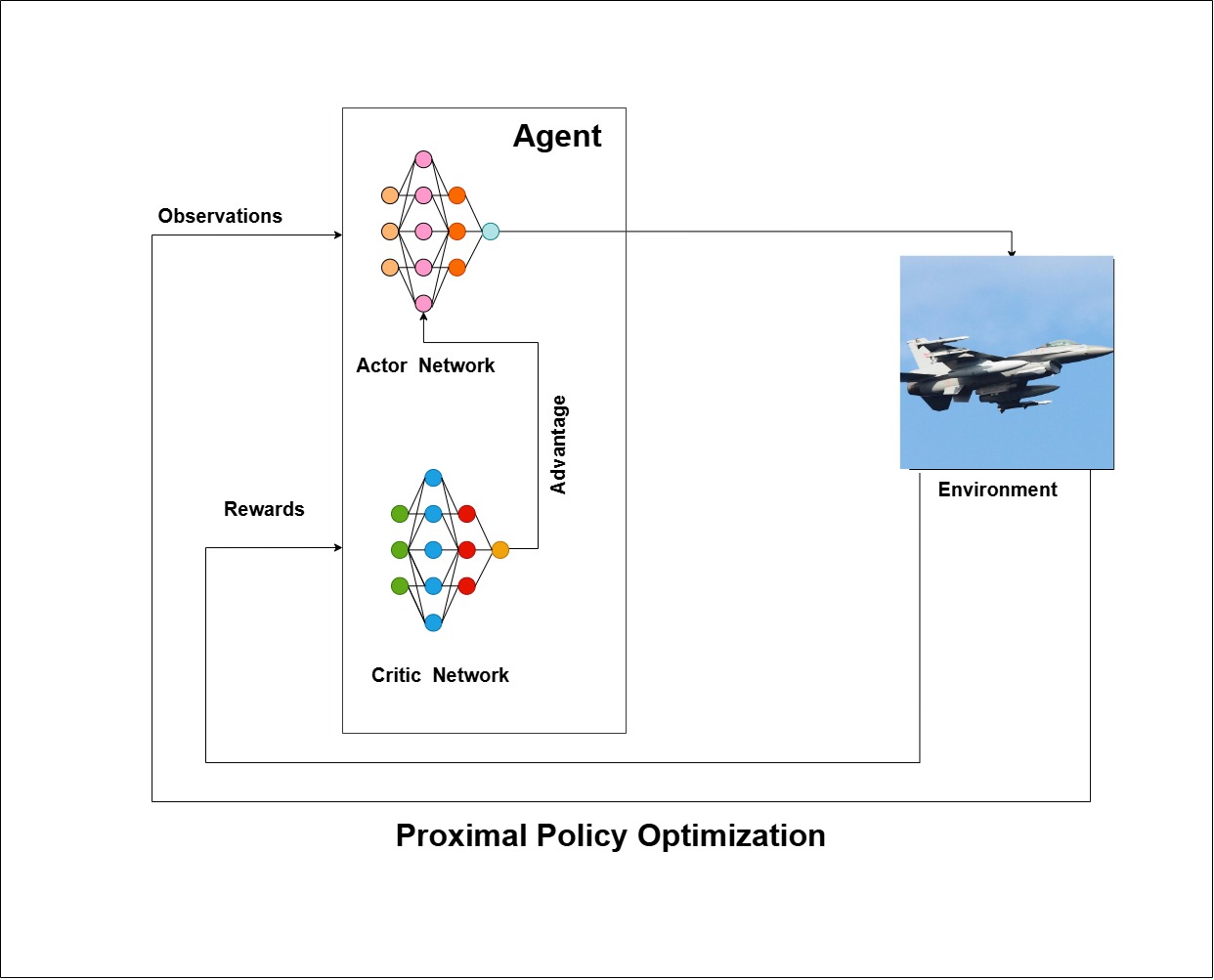}
    \caption{Proximal Policy Optimization Structure for Flight Control.}
    \label{fig:PPO_diagram}
\end{figure}

This work casts the control problem as a Markov Decision Process described by $(s,a,p,r,\Gamma)$: the state, action, transition probability, reward, and a discount factor that balances near-term and long-term returns. PPO constrains how far the new policy may move from the old one within this MDP setting by operating on an on-policy batch and restricting the effective step size. In this work, a reinforcement learning (RL) environment for the aircraft model is developed and trained using Proximal Policy Optimization (PPO). PPO is chosen over off-policy alternatives such as SAC or TD3 for three reasons specific to this application. First, PPO's clipped surrogate objective directly bounds the per-update policy change, which produces smoother, more monotonic reward improvement, a desirable property when the domain-randomized aircraft dynamics and the two-phase reward structure of Section~IV-A already introduce significant training variance. Second, PPO is on-policy and therefore does not rely on a replay buffer of past transitions; this avoids the additional risk, present in off-policy methods, of the critic being updated on stale transitions collected under substantially different aerodynamic parameter samples during domain randomization. Third, PPO's stability and reduced sensitivity to hyperparameter choices, relative to SAC/TD3, is particularly relevant here, since the RL output is subsequently reinterpreted as a bounded nominal compensation term feeding into the SMC-based hybrid architecture (Section~V), where predictable, non-erratic policy updates simplify the characterization of the mismatch bound $\bar\delta$ in Assumption~1. These considerations, rather than raw sample efficiency alone, motivate the choice of PPO over off-policy actor-critic algorithms for this architecture. The architecture of the agent–environment interaction is illustrated schematically in Fig. \ref{fig:PPO_diagram}.

PPO uses a objective function that penalizes the updates of the policy if it updates the policy beyond a threshold value. The objective function is defined by :
\begin{equation}
r_t(\theta) = \frac{\pi_\theta(a_t \mid \bar{s}_t)}{\pi_{\theta_{\text{old}}}(a_t \mid \bar{s}_t)}
\end{equation}

\begin{equation}
    L^{clipped} = E [ min (R(\theta). A_t, clip(R(\theta), 1- \epsilon, 1 + \epsilon)A_t )] .
\end{equation}

Here $E$ is Expectation, $\epsilon$ is the clipping range, $A_t$ represents the advantage value.

The algorithm of PPO can be described as:

\begin{center}
\small
\centering
\begin{tabular}{@{}l@{~}p{7 cm}@{}}
\hline
\textbf{Algorithm 1} & \textbf{PPO algorithm} \\
\hline
1:  & Initialize the actor network $\pi_\theta$ and the value network $V_\phi$ \\
2:  & \textbf{for} Iteration $= 1, 2, \dots, M$ \textbf{do} \\
3:  & \hspace{1em} \textbf{for} Episode $= 1, 2, \dots, N$ \textbf{do} \\
4:  & \hspace{2em} \textbf{for} $t = 1, 2, \dots, T$ \textbf{do} \\
5:  & \hspace{3em} Using policy $\pi_\theta$ choose an action $a_t$ \\
6:  & \hspace{3em} If termination conditions hold true, end the episode \\
7:  & \hspace{2em} \textbf{end for} \\
8:  & \hspace{1em} Find advantage values ${A}_1, {A}_2, \dots, {A}_T$ using $V_\phi$ \\
9:  & \hspace{1em} \textbf{end for} \\
10: & \hspace{1em} \textbf{for} Epochs $= 1, 2, \dots, K$ \textbf{do} \\
11: & \hspace{2em} Update actor network using value $L^{\text{CLIP}}$ \\
12: & \hspace{2em} Update value using critic loss\\
13: & \hspace{1em} \textbf{end for} \\
14: & \textbf{end for} \\
\hline
\end{tabular}
\end{center}

\subsection{Sliding Mode Control}
Because of their intrinsic non-linearity and discontinuity, Sliding Mode Controllers (SMC) are particularly well-suited for the control of intricate non-linear systems such as aerial vehicles. In SMC, the approach includes the specification of a sliding surface, and an appropriate switching logic is formulated to keep the system trajectories on the surface. When the trajectories stray from the surface, they are pushed back onto the sliding surface. This corrective action, however, causes small-amplitude, high-frequency oscillations in the system inputs, an effect known as chattering [1][2]. Many techniques can be used to reduce this chattering, such as higher-order sliding mode control [3][4], boundary layer techniques, and the power rate reaching law \cite{r28,r32}.

Sustaining a leveled flight state requires state controls $\alpha$, $\beta$, and $\mu$. These are the angle of attack ($\alpha$) of the aircraft, angle of sideslip ($\beta$), and wind-axis roll angle ($\mu$). Each variable is assigned to stability in one axis: $\alpha$ for pitch, $\beta$ for yaw, and $\mu$ for roll. The controller omits the control surface aerodynamic force terms compared to control moments because their influence is negligible \cite{r29}. This simplification also removes non-minimum phase behavior from the controller dynamics, which facilitates an easier control law design \cite{r30, 10927436}. Under this assumption, it can be seen from the \eqref{eq:vel} to \eqref{eq:mu} that for each output variable, a control input term appears only after the output variable is differentiated twice with respect to time. This sets the relative degree ($\varsigma$) of each of the output variables to 2. The relative degree of a state is the number of its time derivatives needed before the input can be seen within the resulting expression. The aggregate relative degree for this system is 6, which is less than the overall system order \cite{r31}. Here, the system order is 8 and ($\varsigma_{sys} = 6 < 8$) minus the combined relative degree represents the internal dynamics of the system. These are shaped by the throttle input, which is maintained at a steady value \cite{r6}. With the help feedback linearization, one can convert the coupled system of order 6 to 3 decoupled systems of order 2 which can be seen in \eqref{eq:linearized model}. Since feedback linearization decouples the sixth-order coupled system into three independent second-order subsystems in the auxiliary input $\nu$, each output channel $y_i \in \{\alpha,\beta,\mu\}$ admits an independent scalar sliding-mode design. For notational simplicity, the sliding-mode controller is therefore first derived below for a single representative channel treating all variables as scalars; the identical construction is applied independently to each of the three decoupled channels, and the composite vector quantities $u$, $s$, and $\mathrm{sign}(s)$ used in the hybrid formulation of Sections IV-C and V are obtained by stacking the three per-channel scalar solutions.

\subsection*{Remark on Internal Dynamics}
Since the controlled outputs $y=[\alpha,\beta,\mu]$ have a combined relative
degree of $6$, and the full system order is $8$, the remaining two states,
airspeed $V$ and flight-path angle $\gamma$, are not directly controlled by
$u$ and form the internal dynamics of the closed-loop system. With the
throttle $\eta$ held fixed, \eqref{eq:vel} and \eqref{eq:mu} show that
$\dot V$ and $\dot\gamma$ depend only on $\alpha$, $\beta$, $\mu$, and
$\gamma$ itself, the same states that the proposed controller already drives
to bounded, stabilized values. Once $\alpha$, $\beta$, $\mu$ and the body
rates $p,q,r$ are stabilized, the right-hand sides of \eqref{eq:vel} and
\eqref{eq:mu} settle to bounded values as well, so $V$ and $\gamma$ also
settle to some finite value rather than growing without bound. This is a
direct consequence of the aircraft's own force-balance equations under fixed
throttle, and does not require any additional control action.

It is important to note that this only shows $V$ and $\gamma$ remain
bounded and settle to \emph{some} finite value, not that they are driven to any specific commanded value. The present controller and reward structure are designed to recover attitude and body rates only; $V$ and $\gamma$ are simply left to settle wherever the internal dynamics take them.

This boundedness argument holds within the normal operating envelope
considered in this work. It is worth noting that $\dot\alpha$ in
\eqref{eq:alpha} itself depends on $V$ through the aerodynamic force term
$\tfrac{1}{mV}\{\cdots\}$; at very low airspeed, or at extreme values of
$\gamma$ near the coordinate singularities discussed in Section~III-B ($V
\to 0$ or $\gamma = \pm 90^\circ$), this coupling term is no longer small and can, in principle, feed back into the $\alpha$ dynamics strongly enough to affect its convergence. Such conditions lie outside the normal flight envelope considered in this work, and the internal-dynamics boundedness argument above should accordingly be understood as holding on this same envelope, a compact operating region formalized later as Assumption~1 in Section~V, rather than as a global claim.

Both regulating $V$ and $\gamma$ to specific reference values (e.g. to hold a target airspeed or flight-path angle after recovery) and keeping the aircraft within the compact envelope at the edge cases noted above could in principle be handled by adding an outer guidance/stabilizing loop on top of the present architecture; however, such a loop is not required for, and is outside the scope of, the present study, whose objective is limited to inner-loop attitude and rate recovery from loss-of-control.

For a second order system ($\varsigma = 2$),

\begin{align}
    \dot{x}_1 = x_2
\end{align}

\begin{align}
    \dot{x}_2 = f(x) + g(x)u 
\end{align}

\begin{align}
    y = h(x)
\end{align}

Let $e = y - y_{ref}$ be the tracking error.

Therefore,

\begin{align}
    \dot{e} = \dot{y} - \dot{y}_{ref}
\end{align}

\begin{align}
    \ddot{e} = \ddot{y} - \ddot{y}_{ref}
\end{align}

Constructing the sliding surface $s$ as:

\begin{align}
    s = \dot{e} + \lambda_1 e + \lambda_2 \int_0 ^t e \:  d \tau \:.
\end{align}

Designing control $u$ as:

\begin{align}
    u = g(x)^{-1}(u_r + u_s) 
\end{align}

\begin{align}
    u_r = -f(x) + \ddot{y}_{ref} -  \lambda_1  \dot{e} + \lambda_2 e
\end{align}

\begin{align}
    u_s = -(K |s|^{1/2} + \epsilon) \, \mathrm{sign}(s) \: ,
\end{align}
where $s$, $K$, and $\epsilon$ are the scalar sliding variable and
gains associated with the representative decoupled channel introduced above.

The output variables are $y = [\alpha, \beta , \mu]$. Since, there are couplings in longitudinal and lateral-directional dynamics of the aircraft, this work uses the input-output linearization to transform the system inputs \cite{r30,r31}.

Consider $z_1 =  y = [\alpha,\beta,\mu]$.

$$
 z_2 =  \dot{y} =\left [ \begin{array}{cc}
   L_f\alpha  &  \\
    L_f{\beta}  & \\
    L_f{\mu} 
\end{array} \right ]
$$

\begin{equation}
\label{eq:linearized model}
\dot{z}_2 = \ddot{y} = \left [ \begin{array}{cc}
   L_f^2 \alpha  &  \\
    L_f^2{\beta}  & \\
    L_f^2{\mu} 
\end{array} \right ]  +  \left [ \begin{array}{cc}
   L_gL_f \alpha  &  \\
    L_gL_f{\beta}  & \\
    L_gL_f{\mu} 
\end{array} \right ]u= \nu
\end{equation}

Here $\nu$ is the auxiliary input and $L_f y $ represents Lie derivative of y along the vector field f(x).

\begin{align}
    L_fy = \frac{\partial y}{\partial x}f(x) = \nabla y^T f(x)
\end{align}

Finally, the control input can be calculated from auxiliary input \cite{r31} as:
$$
u = {L_g L_f y}^{-1}(\nu - L_f^2 y) \: .
$$

The invertibility of $L_gL_f y$ was examined by symbolically evaluating the determinant of $L_gL_f y$. Singularities arise primarily under conditions such as zero airspeed ($V = 0$) or coordinate singularities of the wind--axis representation (e.g., $\beta = \pm 90^{\circ}$ or $\gamma = \pm 90^{\circ}$). Since these conditions lie outside the normal flight envelope, $L_gL_f y$ remains non-singular in the operating region, ensuring that the control law is well defined.

\subsection{RL-SMC based Hybrid control}

In this work, hybrid control design that pairs a reinforcement-learning module with a sliding-mode feedback law was developed. The reinforcement-learning policy is trained in simulation across a wide range of starting states, aerodynamic parameters, sensor noise, delays, and gusts. After training, it runs onboard as a fast state-to-command mapping that proposes anticipatory, coordinated surface deflections. In practice, it learns the kinds of “look-ahead” moves that shorten transients, reduce overshoot, and take advantage of natural couplings between roll, pitch, and yaw. Fig. \ref{fig:Block_Diagram3} shows the proposed architecture's block diagram.

Because a learned policy does not come with formal guarantees and can misgeneralize outside its training envelope, its contribution is deliberately limited and checked before it reaches the actuators. It is important to place a hard bound on how much authority the learned policy may exercise and pass its suggestion through a safety filter. This filter enforces deflection and rate limits, honors actuator dynamics, and ensures the baseline stability condition is not compromised. If the learned suggestion would violate these rules, the filter minimally modifies it; if necessary, it blocks it entirely.

The reinforcement-learning policy is used as a learned nominal control component that captures coordinated nonlinear behavior from interaction with the aircraft model. In the proposed architecture, the RL policy primarily generates the maneuvering control effort and cross-axis coordination, while the sliding-mode controller acts as a robust corrective feedback term that compensates disturbances, uncertainty, and residual mismatch. Consequently, the RL pathway is interpreted as a learned nominal compensation channel for nominal dynamics generation, whereas stability and robustness are ensured through the SMC feedback loop.

The backbone of the system is the sliding-mode controller. It is responsible for meeting the tracking goal and for robustness against modeling errors and disturbances that act through the same channels as the control surfaces. The sliding-mode law combines a steady equivalent component that cancels the known part of the dynamics with a high-gain corrective component that guarantees fast convergence into a small neighborhood of the target. Its gains are selected to dominate the worst-case combination of plant disturbance and the maximum contribution allowed from the learned policy. In effect, the filtered learned input appears to the sliding-mode analysis as a bounded, matched perturbation that has already been accounted for.

Training of the learned policy reflects these design choices. Rewards emphasize quick damping of body-rate excursions, prompt but well-behaved attitude recovery, low control activity changes from one instant to the next, and strict compliance with deflection and rate limits. Domain randomization ensures the policy sees many plausible aircraft conditions rather than a single “perfect” model. At deployment, the allowed authority for the learned policy can be scheduled with flight condition: higher in well-modeled regions, lower near stall, saturation, or other edges of the envelope.

This arrangement preserves the strengths of both components. The sliding-mode controller provides the certificate of stability and robustness; the learned policy improves performance within clearly defined, enforced limits. A supervisory layer monitors out-of-distribution states and repeated safety interventions; when these are detected, the system falls back to sliding-mode-only operation until conditions return to normal. The result is a controller that keeps the guarantees expected in flight-critical applications while harvesting the performance gains that modern learning methods can offer.

\section{Implementation}

As discussed in section II, first an RL based environment for the aircraft model is created using OpenAI gym. The Observation space and action space of the gym environment is defined as follows:

\begin{table}[ht]
\footnotesize
\centering
\caption{Observation and Action Space for F-18 RL Environment (All Angles in Radians)}
\label{tab:obs_action_space}
\begin{tabular}{|c|p{2.5cm}|c|c|c|}
\hline
\textbf{Type} & \textbf{Variable (Unit)} & \textbf{Dim.} & \textbf{Range} & \textbf{Ref.} \\
\hline
Obs. 
  & Airspeed $V$ (ft/sec) & 1 & $[0, 2000]$ & — \\
  & AoA $\alpha$ (rad) & 1 & $[-0.244, 1.571]$ & $\alpha_{ref}$ \\
  & Sideslip $\beta$ (rad) & 1 & $[-\pi, \pi]$ & $\beta_{ref}$ \\
  & Bank angle $\mu$ (rad) & 1 & $[-\pi, \pi]$ & $\mu_{ref}$ \\
  & Angular rates $p, q, r$ (rad/s) & 3 & $[-10\pi, 10\pi]$ & $0$ \\
  & Flight path angle $\gamma$ (rad) & 1 & $[-1.745, 1.745]$ & — \\
\hline
Act. 
  & Elevator $\delta_e$ (rad) & 1 & $[-0.436, 0.175]$ & — \\
  & Aileron $\delta_a$ (rad) & 1 & $[-0.436, 0.436]$ & — \\
  & Rudder $\delta_r$ (rad) & 1 & $[-0.524, 0.524]$ & — \\
\hline
\end{tabular}
\normalsize
\end{table}

\subsection{Training of the RL Model}

To effectively control the aircraft and to bring it to a steady flight condition, the angular rates ($p$ , $q$ and $r$) of aircraft should be brought down to lower values; after this to maintain a leveled flight condition, the control of states $\alpha$, $\beta$, and $\mu$ is required which helps in attitude stabilization. From \eqref{eq:vel} and \eqref{eq:mu}, it can seen that as long as $\alpha$, $\beta$, $\mu$ and angular rates are stabilized, $V$ and $\gamma$ will also be stabilized to finite value, subjected to condition that $\eta$ does not change.

The first priority should be attenuation of the angular rates and then create a leveled flight attitude condition. Therefore, the design of the reward function should be carefully done so that it emphasizes more on controlling the angular rates initially, and once the angular rates drop below a threshold value, then the reward function should prioritize maintaining the leveled flight at target angles \cite{doi:10.2514/1.G001739}. Designing a continuous reward structure for the RL environment is done in two phases. In the first phase, agent only focuses on rates and in the second phase, agent prioritizes attitude angles in addition to angular rates. To accelerate the training, the Potential Based Reward Shaping (PBRS) is also used \cite{Ng1999PBRS}.

The reward structure is defined as:

\textit{Phase 1}

When $[p,q,r] > 0.17 \: rad/sec$
\begin{align}
    r_{phase_1} = -|| \omega ||^{2} - w_{p_1} ( |pq| + |qr| + |pr|) \: .
\end{align}

Where:

\begin{align*}
    ||\omega ||  = \sqrt{p^2 + q^2 + r^2} \: .
\end{align*}

$|pq| + |qr| + |pr|$ term represents the couplings between the three axes. These product terms come directly from the rotational dynamics in \eqref{eq:p}--\eqref{eq:r} (for example, the $\tfrac{I_y-I_z}{I_x}qr$ term in the equation for $\dot p$), and they control how much rotation in one axis spills over into the other two. During agile maneuvers such as spin recovery, the body rates are large, so this spillover is significant and motion is being transferred from the dominant spin axis into the other axes instead of all three rates settling down together. Penalizing $|pq|+|qr|+|pr|$ during Phase~1 discourages the agent from relying on this transfer and instead pushes it toward reducing all three body rates at the same time. This penalty is added for the agile maneuvers, not a general claim that axis coupling is always destabilizing. $w_{p_{1}}$ is the weight factor for scaling the weights of the rate penalty and cross-coupling penalty.

\vspace{5 pt}

\textit{Phase 2}

\vspace{5 pt}

When $||\omega ||  < 0.17 \: rad/sec$
\begin{align}
    r_{phase_2} &= \ -e_{\alpha}^{2} - w_{p_{21}}e_{\alpha} q  \notag \\
    & \ \ \ \ - w_{p_{22}}(|| \omega ||^{2} +  |pq| + |qr| + |pr|) 
\end{align}

\vspace{15 pt}

Additional Reward:
           \begin{align*}
               & \text{+5 if } |e_\alpha| < 0.017 \ \ rad \\
               & \text{+3 if } |e_\beta| < 0.017 \ \ rad \\
               & \text{+3 if } |e_\mu| < 0.017 \ \ rad \\
               & - \sum_i action_{i}^2  \: .
           \end{align*}

\vspace{15 pt}

Where:

\begin{align*}
    e_{\alpha} &= \alpha \: - \alpha_{d} \\
    e_{\beta} &= \beta \: - \beta_{d} \\
    e_{\mu} &= \mu \: - \mu_{d} \:.
\end{align*}

A sudden jump in reward change from phase 1 to phase 2 may affect the training so a logistic gate blend function $\varepsilon(||w||)$ is added to make a smooth shift from phase 1 to phase 2.

Let
\begin{equation}
  \varepsilon \!\left(\lVert \omega \rVert\right)
  = \frac{1}{1 + \exp\!\big(k_{\mathrm{blend}}\big(\lVert \omega \rVert - \omega_{\mathrm{th}}\big)\big)},
\end{equation}
with \(k_{\mathrm{blend}} = 10\) and \(\omega_{\mathrm{th}} = 0.17 rad /\mathrm{s}\). The raw performance reward is
\begin{equation}
  r_{\mathrm{raw}} = (1 - \varepsilon)\, r_{\mathrm{phase1}} + \varepsilon\, r_{\mathrm{phase2}} \label{eq:r_raw} \:.
\end{equation}
When rates are high, \(\varepsilon \approx 0\) and Phase~1 dominates; when rates are small, \(\varepsilon \approx 1\) and Phase~2 dominates.

The variables \( \alpha_d \), \( \beta_d \), and \( \mu_d \) denote the target values for the angle of attack, sideslip angle, and bank angle, respectively. To simplify the design of the reward structure, continuous rewards are not explicitly assigned for deviations in sideslip angle \( \beta \) and bank angle \( \mu \). This is justified by the inherent symmetry of the aircraft, which suggests that these states will naturally tend toward zero over time. Instead, the stabilization of roll and yaw rates \( p \) and \( r \) is expected to indirectly contribute to restoring \( \beta \) and \( \mu \) to their nominal values. In the second reward phase, the weighting coefficient \( w_{p_{22}} \) should be chosen carefully—large enough to ensure correction of coupling errors and rate stabilization, but not so large that it overwhelms the other objectives.

Traditional continuous penalty-based rewards apply a constant cost at each step, regardless of whether the agent is progressing toward or away from the goal. In contrast, Potential-Based Reward Shaping (PBRS) introduces direction-sensitive feedback, assigning positive values when the agent moves closer to the desired state and negative values when it deviates \cite{Ng1999PBRS}. Although PBRS does not alter the set of optimal solutions, it significantly improves training efficiency by offering a more informative gradient to guide policy learning.

\begin{align}
    \sigma = -e_{\alpha}^2 - w_{p_{21}}||\omega||^2 \label{eq:pbrs}
\end{align}

Here $\sigma$ is the reward shaping potential.
From \eqref{eq:r_raw} and \eqref{eq:pbrs}, the total reward achieved is:

\begin{align}
    \text{Reward} = r_{raw} + \Gamma  \sigma_{t+1} - \sigma_{t} \:.
\end{align}

The term \( \sigma_t \) represents the potential function evaluated at the current timestep, while \( \sigma_{t+1} \) corresponds to the potential at the subsequent step, based on the predicted trajectory. The discount factor \( \Gamma \) is a critical hyperparameter in PPO training, controlling how much the agent values long-term rewards relative to immediate ones. This same discount factor is also employed in shaping the PBRS-based reward, ensuring consistency between policy optimization and reward guidance.

For training and evaluation, the nonlinear dynamics of the F-18/HARV aircraft model were implemented within a custom environment based on the OpenAI Gym interface \cite{Brockman2016Gym}. This tailored Gym environment was designed to accurately capture the system's state representation, permissible action set, and underlying equations of motion. Specific terminal conditions were introduced to detect situations where the aircraft exceeds its operational flight envelope or significantly diverges from the desired target states. A large negative reward of \(-1000\) was assigned when such terminal conditions were triggered. This value is decided relative to maximum possible cumulative reward. The structure and control flow governing the environment's simulation logic are described in Algorithm~2. Detailed bounds for the observation and action spaces are provided in Table~\ref{tab:obs_action_space}.

\begin{center}
\small
\centering
\begin{tabular}{@{}l@{~}p{7 cm}@{}}
\hline
\textbf{Algorithm 2} & \textbf{F-18 Environment Logic 6 DOF} \label{algo:f8_env}\\
\hline
1:  & \textbf{Input:} Initial state $x_0$, timestep $\Delta t$, control $a_t$ \\
2:  & Initialize F-18 state vector and aircraft parameters \\
3:  & \textbf{function} \texttt{reset()} \\
4:  & \hspace{1em} Sample target angle of attack $\alpha \in [-5^\circ, 40^\circ]$ \\
5:  & \hspace{1em} Initialize state $x_0 = [V, \alpha, \beta, p, q, r, \dots]$ \\
6:  & \hspace{1em} \textbf{return} initial observation $\bar{s}_0$ \\
7:  & \textbf{end function} \\
8:  & \textbf{function} \texttt{step}($a_t$) \\
9:  & \hspace{1em} Clip and scale $a_t$ to control limits \\
10: & \hspace{1em} Compute control input $u_t$ using Env actions $a_t$\\
11: & \hspace{1em} Integrate aircraft dynamics: $x_{t+1} = f(x_t,u_t)$ \\
12: & \hspace{1em} Update positions and Euler angles \\
13: & \hspace{1em} Compute reward: Potential-based reward shaping or error-based penalty \\
14: & \hspace{1em} Check termination conditions \\
15: & \hspace{1em} \textbf{return} $(\bar{s}_{t+1}, r_t, d_t, \text{info})$ \\
16: & \textbf{end function} \\
\hline
\end{tabular}
\end{center}

The observation vector consists of the following states:

\begin{align}
    \bar{s} = \{ V, \alpha, \beta, p, q, r, \mu, \gamma, \alpha_{d}, \beta_{d}, \mu_{d} \}
\end{align}

Action space is defined as:

\begin{align}
    a = \{\delta_e , \delta_a , \delta_r \}
\end{align}

Simulation is randomly initialized at the random state in the flight envelope at each episode. The bounds of the random states for episode initialization are shown in Table \ref{tab:f18_state_bounds_rad}. At the start of simulation the altitude of aircraft is assumed at $0$ ft. so that loss of altitude can be analyzed with ease.

\begin{table}[ht]
\footnotesize
\centering
\caption{Bounds for random initialization of state (Angles in Radians; Speed in ft/sec)}
\begin{tabular}{|p{3.6 cm}|c|c|}
\hline
\textbf{State Variable (Unit)} & \textbf{Lower Bound} & \textbf{Upper Bound} \\
\hline
Airspeed $V$ (ft/sec) & 100 & 1500 \\
Angle of Attack $\alpha$ (rad) & $-0.175$ & $1.047$ \\
Sideslip Angle $\beta$ (rad) & $-0.262$ & $0.262$ \\
Roll Rate $p$ (rad/s) & $-2.094$ & $2.094$ \\
Pitch Rate $q$ (rad/s) & $-1.047$ & $1.047$ \\
Yaw Rate $r$ (rad/s) & $-0.524$ & $0.524$ \\
Bank Angle $\mu$ (rad) & $-3.142$ & $3.142$ \\
Flight Path Angle $\gamma$ (rad) & $-1.047$ & $1.047$ \\
Desired AoA $\alpha_d$ (rad) & $-0.085$ & $0.68$ \\
Desired Sideslip $\beta_d$ (rad) & 0 & 0 \\
Desired Bank Angle $\mu_d$ (rad) & 0 & 0 \\
\hline
\end{tabular}
\normalsize
\label{tab:f18_state_bounds_rad}
\end{table}

The weights $w_{p_1}$, $w_{p_{21}}$ and  $w_{p_{22}}$ were kept 0.3, 0.3 and 0.05 respectively. Target value $\alpha_d $ is also randomized for every episode between -0.085 radians to 0.68 radians. $\beta_d$ and $\mu_d $ are kept to zero. The control inputs are subject to predefined saturation limits to simulate real-world actuator constraints. The actuator constraints for elevator, aileron and rudder are ($-25^o$,\:$10^o$), ($-25^o$,\:$25^o$) and ($-30^o$,\:$30^o$) respectively. The control objective is to track reference trajectories for key state variables (e.g., $\alpha$, $p$, $q$, $r$ etc.) while ensuring that the system can converge to the desired values. 
Training of the RL agent was carried out using Stable-Baselines3(SB3) which is a popular Python-based open-source library for reinforcement learning algorithms \cite{Raffin2021SB3}. Proximal Policy Optimization Algorithm was used as it is sample efficient and robust in the training process. The network architecture is an Actor-Critic Network of (256, 128) hidden layers - Tanh configuration each. This size was decided empirically after running multiple simulations with different network sizes to avoid overfitting. Learning rate and Discount factor were kept at 5e-5 and 0.99 to encourage smooth learning and long-term rewards. The agent was trained for a total of 6000 episodes. Each episode consists of 20000 timesteps, with one timestep being of size 0.01 seconds. Agent training performance was logged to check the reward trend with ongoing training using Tensorboard. The trained model will give the output $u_{RL}$ which is treated as the learned nominal compensation term $ u_{\text{ff}}$.

\subsection{Integration of control architectures}

For integrating the nominal compensation term in the SMC feedback controller, some modifications to the system formulation are required.

\begin{align}
    \dot{x}_1 &= x_2 \\
    \dot{x}_2 &= f(x) + g(x) u + d(x,t) 
\end{align}

The scalar $\zeta \in (0,1]$ is introduced as a {control authority factor} that modulates the contribution of the sliding mode feedback term. Specifically, $\zeta$ determines the relative dominance of the feedback controller with respect to the RL-based policy.

Now, the control input is a combination of learned feedforward and feedback terms:
\begin{align}
    u =  u_{\text{ff}} +  u_{\text{fb}} = u_{\text{RL}} + \zeta u_{\text{SMC}} \label{eq:total_input}
\end{align}

Without the introduction of the scaling factor $\zeta$, the high-gain and discontinuous nature of the sliding mode controller can dominate the composite control input, thereby suppressing the behavior learned by the reinforcement learning policy. The parameter $\zeta$ therefore enables a structured blending of learning-based control with model-based robust feedback, preserving the performance advantages of the RL control while ensuring stability and disturbance rejection through SMC.

For ideal nominal compensation using only \( u =  u_{\text{ff}} \):
\begin{align}
    \dot{x}_2 = f(x) + g(x)  u_{\text{ff}}
\end{align}

However, if \(  u_{\text{ff}} \) is insufficient for perfect tracking:
\begin{align}
    \dot{\bar{x}}_2 = f(x) + g(x)  u_{\text{ff}} \neq \dot{x}_2 
\end{align}

The resulting mismatch term is:
\begin{align}
    \delta(x)= \dot{\bar{x}}_2 - \dot{x}_2 
\end{align}

Substituting \eqref{eq:total_input} into  the original system dynamics:
\begin{align*}
    \dot{x}_2 = f(x) + g(x) ( u_{\text{ff}} +  u_{\text{fb}}) 
\end{align*}

\begin{align}
    \dot{x}_2 = \dot{\bar{x}}_2 + g(x)  u_{\text{fb}} \label{eq:fb_dynamics}
\end{align}

\eqref{eq:fb_dynamics} defines a new control system for the feedback controller to handle. Therefore, a sliding mode controller (SMC) is designed for the feedback input \(  u_{\text{fb}} \).

Define the tracking error as \( e = x_1 - y_{\mathrm{ref}} \), and the sliding surface as:
\begin{align}
    s = \dot{e} + \lambda_1 e + \lambda_2 \int_0^t e(\tau)\, d\tau
\end{align}

Then, the feedback control input is:
\begin{align}
     u_{\text{fb}} = g(x)^{-1} \left( u_r + u_s \right)
\end{align}

where
\begin{align}
    u_r &= -\dot{\bar{x}}_2 + \ddot{y}_{\mathrm{ref}} - \lambda_1 \dot{e} - \lambda_2 e \label{eq:u_r} \\
    u_s &= -K |s|^{1/2} \mathrm{sign}(s) \label{eq:u_s}
\end{align}

Here $s = [s_1,s_2,s_3]^\top \in
\mathbb{R}^3$ denotes the composite sliding surface obtained by stacking the
three per-channel scalar surfaces associated with
$y=[\alpha,\beta,\mu]$, and $|s|^{1/2}$,
$\mathrm{sign}(s)$ are applied element-wise
(i.e.\ $\mathrm{sign}(s) =
[\mathrm{sign}(s_1),\mathrm{sign}(s_2),\mathrm{sign}(s_3)]^\top$),
so that the product $K|s|^{1/2}\,\mathrm{sign}(s)$ is
understood component-wise.

Finally, after substituting from \eqref{eq:u_r} and \eqref{eq:u_s} the total control input becomes:
\begin{align}
       u =  u_{\text{ff}} + g(x)^{-1} (-\dot{\bar{x}}_2 + \ddot{y}_{\mathrm{ref}} - \lambda_1 \dot{e} - \lambda_2 e \\ \notag - K |s|^{1/2} \mathrm{sign}(s))
\end{align}

\section{Stability Analysis}

Define the nominal compensation mismatch error as:
\begin{align}
    \delta(x) = \dot{\bar{x}}_2 - \dot{x}_2
\end{align}

This measures how far the RL-predicted acceleration \( \hat{\ddot{x}} \) is from the true (ideal) model-based acceleration:

\begin{align}
    \dot{x}_2 = f(x) + g(x) u_{\text{ff}}^{\text{ideal}} + d(x,t)
\end{align}

The sliding surface is defined as:

\begin{align}
    s = \dot{e} + \lambda_1 e + \lambda_2 \int_0^t e(\tau)\, d\tau \label{eq:sliding_s}
\end{align}

\begin{align}
u = u_{\text{ff}} + u_{\text{fb}}
\end{align}

Differentiating \eqref{eq:sliding_s}:
\[
\dot{s} = f(x) + g(x) u_{\text{ff}} + g(x) u_{\text{fb}} - \ddot{y}_{\text{ref}} + \lambda_1 \dot{e} + \lambda_2 e + d(x,t)
\]

Substitute from \eqref{eq:total_input}:
\begin{align}
    \dot{s} = \dot{\bar{x}}_2 - \ddot{y}_{\text{ref}} + \lambda_1 \dot{e} + \lambda_2 e + g(x) \zeta u_{\text{SMC}}.
\end{align}

Then:
\begin{align}
    \dot{s} = -\zeta\kappa \cdot \text{sign}(s) + \delta(x) + d(x,t).
\end{align}

Where

\begin{align}
    \kappa  = K|s|^{1/2} + \epsilon .
\end{align}

 $\epsilon$ is small positive value that works as a gain when $s \rightarrow 0$
The ideal sliding condition is:
\begin{align}
    \dot{s} = -\zeta\kappa \cdot \text{sign}(s).
\end{align}

But due to imperfect compensation:

\begin{align}
    \dot{s} = -\zeta\kappa \cdot \text{sign}(s) + \delta(x) + d(x,t).
\end{align}

So the term \( \delta(x) + d(x,t) \) acts like a disturbance or uncertainty.

\text{Sliding dynamics:}
\begin{align}
    \dot{s} = -\zeta\kappa \cdot \text{sign}(s) + \delta(x) + d(x,t).
\end{align}

\subsection*{Stability Condition}
\subsection{Lyapunov Stability Analysis}

The proposed control architecture combines a learning-based nominal control policy with a robust Sliding Mode Controller (SMC). The total control input is expressed as

\begin{equation}
u = u_{\mathrm{RL}} + \zeta u_{\mathrm{SMC}},
\label{eq:control_decomposition}
\end{equation}

where $u_{\mathrm{RL}}$ is generated by the reinforcement learning policy, $u_{\mathrm{SMC}}$ denotes the sliding mode feedback controller, and $\zeta>0$ is the control authority factor that determines the contribution of the feedback controller.

Unlike conventional end-to-end learning approaches, the reinforcement learning policy is employed solely to provide nominal performance enhancement, while the sliding mode controller is responsible for guaranteeing robustness and closed-loop stability.

\subsubsection*{Assumption 1}

Consider the nonlinear system

\begin{equation}
\dot{x}=f(x)+g(x)u+d(x,t),
\label{eq:system}
\end{equation}

where the disturbance satisfies

\begin{equation}
\|d(x,t)\|\le \bar d.
\end{equation}

Furthermore, the operating region is assumed to be compact so that

\begin{equation}
\|g(x)\|\le \bar g .
\end{equation}

The reinforcement learning policy is constrained by the actuator limits and the proposed safety filter, yielding

\begin{equation}
\|u_{\mathrm{RL}}\|\le \bar u_{\mathrm{RL}}.
\end{equation}

Assume that there exists an ideal nominal compensation
$u_{\mathrm{RL}}^{*}$ satisfying

\begin{equation}
\|u_{\mathrm{RL}}^{*}\|\le \bar u^{*}.
\end{equation}

Define the learning approximation mismatch as

\begin{equation}
\delta(x)
=
g(x)
\left(
u_{\mathrm{RL}}
-
u_{\mathrm{RL}}^{*}
\right).
\label{eq:delta_definition}
\end{equation}

Then,

\begin{equation}
\|\delta(x)\|
\le
\bar g
\left(
\bar u_{\mathrm{RL}}
+
\bar u^{*}
\right)
=
\bar\delta.
\label{eq:delta_bound}
\end{equation}

Hence, the learning error enters the system as a bounded matched uncertainty.

\begin{theorem}
Consider the nonlinear system (\ref{eq:system}) controlled by (\ref{eq:control_decomposition}). Suppose Assumption 1 holds and the sliding mode controller is designed such that the sliding variable $s$ satisfies

\begin{equation}
\dot{s}
=
-\zeta
\left(
K|s|^{1/2}
+\epsilon
\right)
\operatorname{sign}(s)
+\delta(x)
+d(x,t),
\label{eq:phi_dynamics}
\end{equation}

where $K>0$ and $\epsilon>0$.

If

\begin{equation}
\boxed{
\zeta\epsilon
>
\bar\delta+\bar d,
}
\label{eq:stability_condition}
\end{equation}

then the sliding variable is uniformly ultimately bounded and converges asymptotically to the sliding manifold.
\end{theorem}

\begin{proof}

Choose the Lyapunov candidate

\begin{equation}
V(s)
=
\frac12
s^{T}s.
\label{eq:lyapunov}
\end{equation}

Differentiating gives

\begin{equation}
\dot V
=
s^{T}\dot{s}.
\end{equation}

Substituting (\ref{eq:phi_dynamics}),

\begin{align}
\dot V
&=
-\zeta K|s|^{3/2}
-\zeta\epsilon |s|
+s^{T}\delta(x)
+s^{T}d(x,t).
\end{align}

Using the Cauchy--Schwarz inequality together with (\ref{eq:delta_bound}),

\begin{align}
s^{T}\delta(x)
&\le
\|s\|
\|\delta(x)\|
\le
\bar\delta |s|,
\\
s^{T}d(x,t)
&\le
\bar d |s|.
\end{align}

Therefore,

\begin{equation}
\dot V
\le
-\zeta K|s|^{3/2}
-
\left(
\zeta\epsilon
-
\bar\delta
-
\bar d
\right)
|s|.
\end{equation}

Whenever condition (\ref{eq:stability_condition}) is satisfied,

\begin{equation}
\dot V<0
\qquad
\forall\;
s\neq0.
\end{equation}

Hence, the Lyapunov function decreases monotonically outside the origin, implying that the sliding variable converges to the sliding manifold while remaining robust against bounded learning approximation errors and external disturbances.

\end{proof}

In practice, due to the presence of bounded learning mismatch $\delta(x)$ and
disturbance $d(x,t)$, exact asymptotic convergence to the sliding surface cannot
be guaranteed. Instead, the sliding variable converges in finite time to a small
invariant neighborhood of the sliding manifold.

From the Lyapunov analysis,

\begin{equation}
\dot{V}
\le
-\zeta K |s|^{3/2}
-
\left(
\zeta\epsilon-\bar{\delta}-\bar{d}
\right)
|s|
\label{eq:lyap_bound}
\end{equation}

where

\begin{equation}
V=\frac{1}{2}s^2.
\end{equation}

If the condition

\begin{equation}
\zeta\epsilon>\bar{\delta}+\bar{d}
\label{eq:stability_condition_restated}
\end{equation}

is satisfied, then

\begin{equation}
\dot{V}<0,
\qquad
\forall\;s\neq0,
\end{equation}

which guarantees finite-time convergence of the sliding variable to a small
invariant neighborhood of the sliding manifold despite bounded uncertainty and
residual mismatch.

The nonlinear reaching term $K|s|^{1/2}$ accelerates convergence when
the tracking error is large, while the constant term $\epsilon$ ensures
robustness in the vicinity of the sliding surface \cite{r9,r11}.

This condition establishes a direct relationship between the admissible learning
mismatch, the disturbance bound, and the scaled SMC gain. In particular, it
highlights the role of the SMC control authority parameter $\zeta$ within the
overall control architecture: increasing $\zeta$ enhances robustness to learning
errors and disturbances, while smaller values of $\zeta$ place greater reliance
on the learned nominal compensation component.

In principle, $\zeta$ can be selected or scheduled in several ways, for
instance through a continuous, error-dependent variation that gradually
increases the SMC authority as the tracking error grows. In this work, however,
$\zeta$ is chosen as a two-valued switching law based directly on the bound
established in~\eqref{eq:stability_condition}: $\zeta = 1$ whenever the tracking
error leaves a prescribed bounded tube around the desired trajectory, and
$\zeta = 0$ otherwise. This renders the composite controller a \emph{tube-based}
hybrid scheme: as long as the state remains within the certified error tube, the
learned policy is allowed to act without SMC correction, since the mismatch and
disturbance terms are, by design, within the bound guaranteed in
\eqref{eq:stability_condition}; the instant the error leaves this tube, the SMC
feedback is switched fully on ($\zeta = 1$) to enforce the stability condition
and pull the trajectory back toward the tube. This switching logic keeps the
sliding-mode feedback dormant during nominal, well-modeled operation, where the
RL policy's anticipatory behavior is preserved untouched, while guaranteeing
that the robust correction is invoked precisely when it is needed, i.e., when
the bounded-mismatch assumption is at risk of being violated.

The width of the error tube is chosen with direct reference to the stability
condition~\eqref{eq:stability_condition}. Within the training envelope, the
learning mismatch and disturbance terms are empirically observed to remain
within the bound $\bar\delta+\bar d$ used in Assumption~1 as long as the
tracking error stays within a neighborhood of the reference trajectory; the
tube boundary is set at this empirically justified error level. Consequently,
operating inside the tube corresponds to the regime in which the bounded-
mismatch assumption is expected to hold and $\zeta=0$ (RL-only authority) does
not violate~\eqref{eq:stability_condition}, whereas exceeding the tube is
treated as a signal that $\delta(x)+d(x,t)$ may approach or exceed
$\bar\delta+\bar d$, at which point $\zeta=1$ is triggered so that the SMC gain
$\epsilon$ re-establishes $\zeta\epsilon>\bar\delta+\bar d$ and pulls the
trajectory back within the certified region.

A single-threshold switching rule of this form can in principle induce
high-frequency switching (Zeno-like behavior) if the tracking error dwells
exactly at the tube boundary, since noise or residual mismatch could then
drive $\zeta$ between $0$ and $1$ arbitrarily fast. To preclude this, the
switching law is implemented with a hysteresis band using two thresholds
$e_{\mathrm{on}} > e_{\mathrm{off}} > 0$ rather than a single boundary: $\zeta$
is switched to $1$ only once the tracking error exceeds $e_{\mathrm{on}}$, and
is switched back to $0$ only once the error has decreased below the strictly
smaller threshold $e_{\mathrm{off}}$. Because the error must traverse the
finite gap $e_{\mathrm{on}}-e_{\mathrm{off}}$ before any further switch can
occur, and the sliding dynamics under either fixed value of $\zeta$ evolve with
bounded rate (Theorem~1), the time between consecutive switching events is
bounded below by a strictly positive dwell time. This rules out infinitely
fast switching (Zeno behavior) at the tube boundary while preserving the
qualitative behavior of the tube-based scheme described above.

It is worth noting that exact finite-time convergence to
$s = 0$ in the presence of bounded disturbances can be achieved by
employing a discontinuous switching law (e.g.,
$K\operatorname{sign}(s)$) with sufficiently large gain. However, such
designs sacrifice the smoothness of the proposed power-rate reaching law and may
introduce chattering, which is undesirable for practical actuator
implementation. Since the primary objective of the proposed learning-augmented
control framework is to guarantee robustness and bounded closed-loop behavior in
the presence of RL approximation errors, exact asymptotic convergence is not
required. Instead, finite-time convergence to a small invariant neighborhood of
the sliding surface is sufficient to ensure safety and reliable performance. This also directly tells us that size of corrective sliding gain depends on the mismatch term $\delta(x)$ directly. If one ignores the external disturbance term $d(x,t)$ for simplicity, then if the RL model is perfectly trained such that $\delta(x) \rightarrow 0$, the boundary layer shrinks to the origin, which means control actions are fully generated by RL agent.

\section{Simulation Results and Discussions}

To evaluate the proposed architecture, this work considers three experimental phases. The first phase presents a Monte Carlo evaluation of the fully trained hybrid controller's robustness to randomized initial flight conditions.

\begin{figure}[H]
    \centering
    \subfloat[Angle of attack $\alpha$]{
        \includegraphics[width=0.98\linewidth]{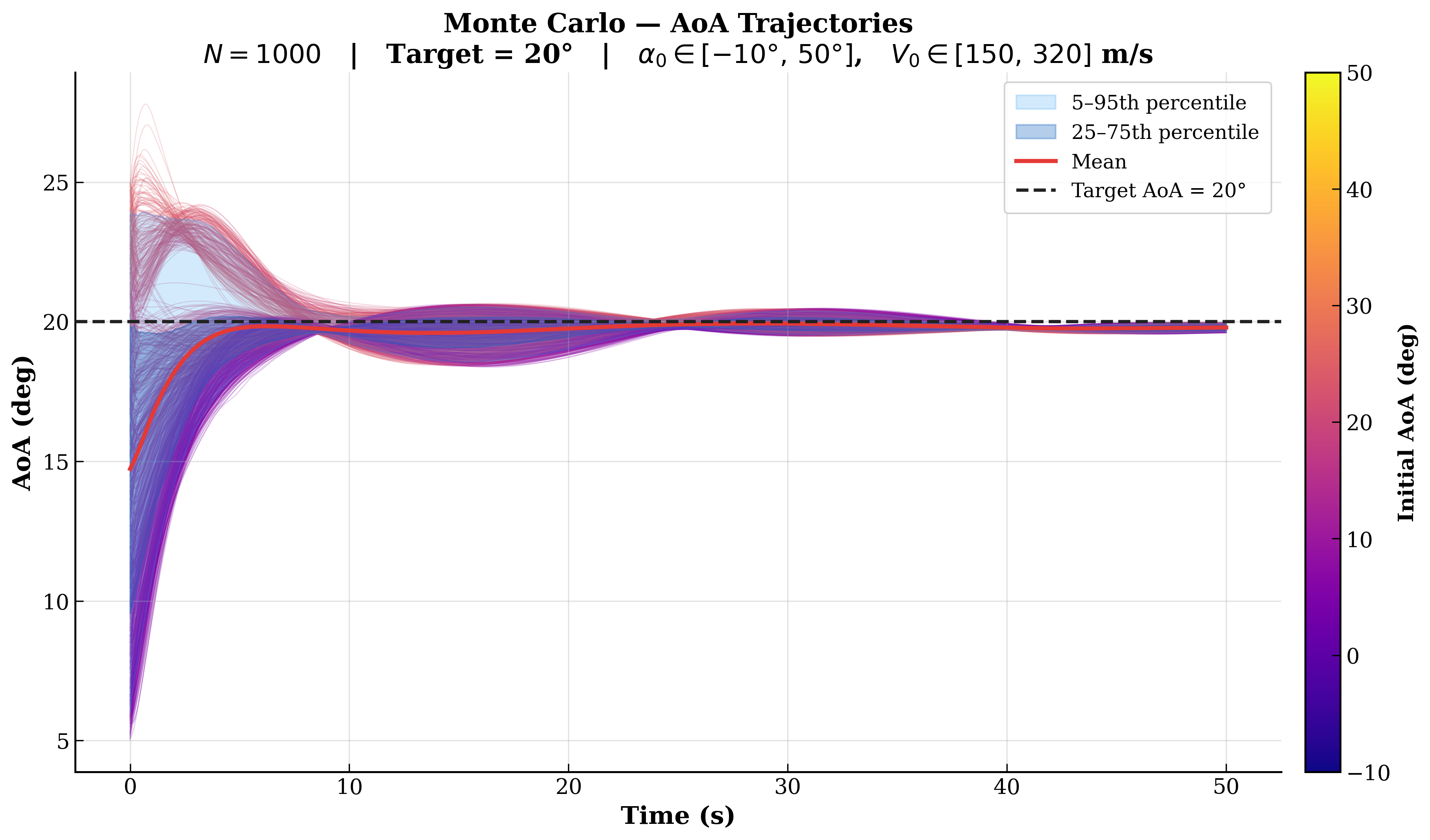}}
    \\
    \subfloat[Sideslip $\beta$]{
        \includegraphics[width=0.98\linewidth]{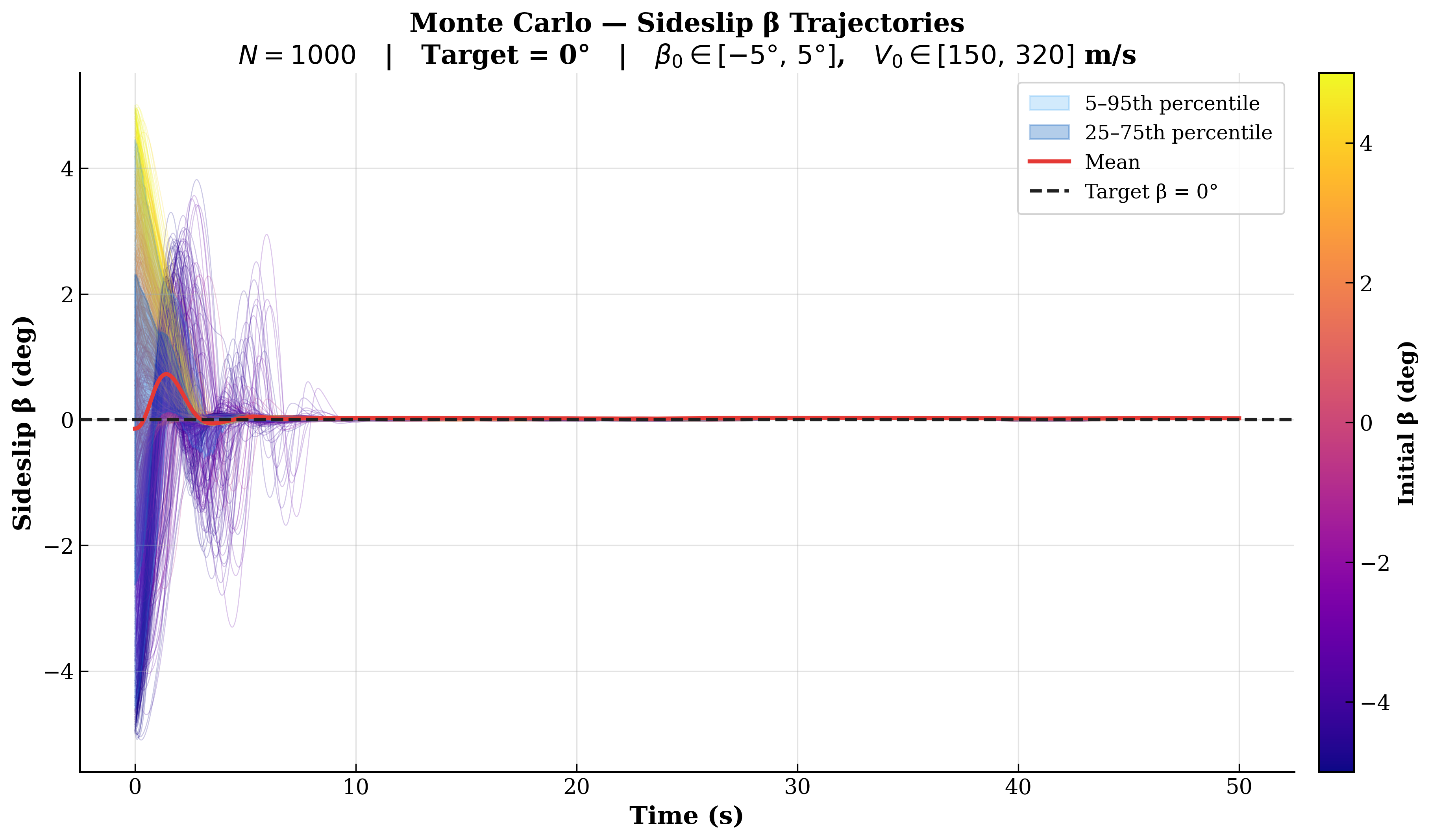}}
    \\
    \subfloat[Bank angle $\mu$]{
        \includegraphics[width=0.98\linewidth]{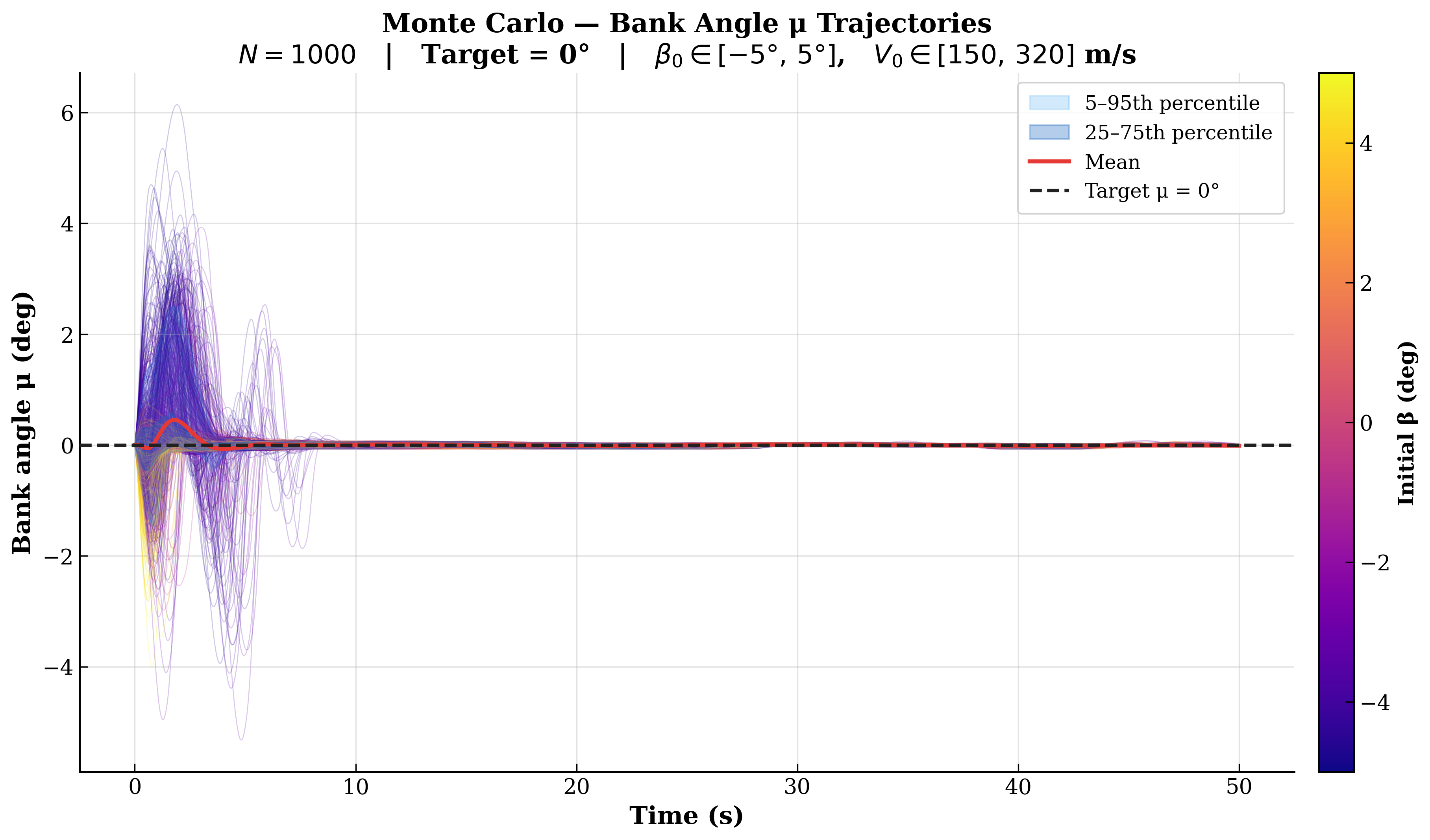}}
    \caption{Monte Carlo trajectory bands ($N=1000$) for the three controlled outputs under randomized initial conditions.}
    \label{fig:mc_trajectories}
\end{figure}

In the second phase, the effect of introducing the SMC safety layer into a {partially trained} RL policy is analyzed. In the third phase, a full comparative study of \emph{RL-only}, \emph{SMC-only}, and \emph{RL--SMC hybrid} controllers using a {fully trained} RL policy tuned for spin recovery is conducted.

\subsection{Monte Carlo Robustness Evaluation}

To assess robustness to variation in initial flight condition, $N=1000$ Monte Carlo trials were run with initial airspeed $V_0$, angle of attack $\alpha_0$, sideslip $\beta_0$, and body rates $p_0,q_0,r_0$ sampled uniformly over the ranges given in Table~\ref{tab:f18_state_bounds_rad}, for the fully trained hybrid RL-SMC controller with a fixed target angle of attack of $20^\circ$  (0.349 radian). 

\begin{figure}[H]
    \centering
    \subfloat[Terminal $\alpha$ error]{
        \includegraphics[width=0.98\linewidth]{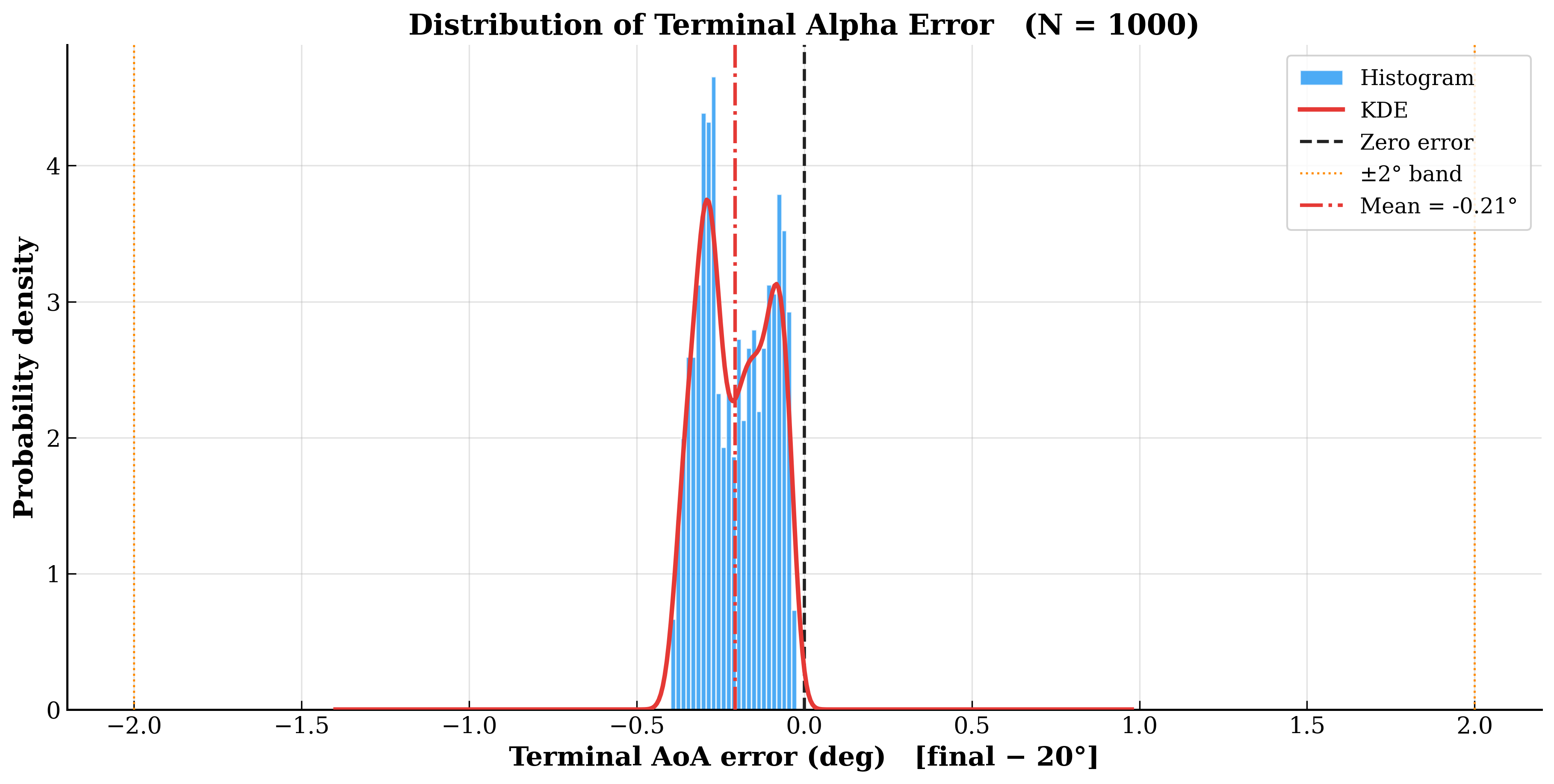}}
    \\
    \subfloat[Terminal $\beta$ error]{
        \includegraphics[width=0.98\linewidth]{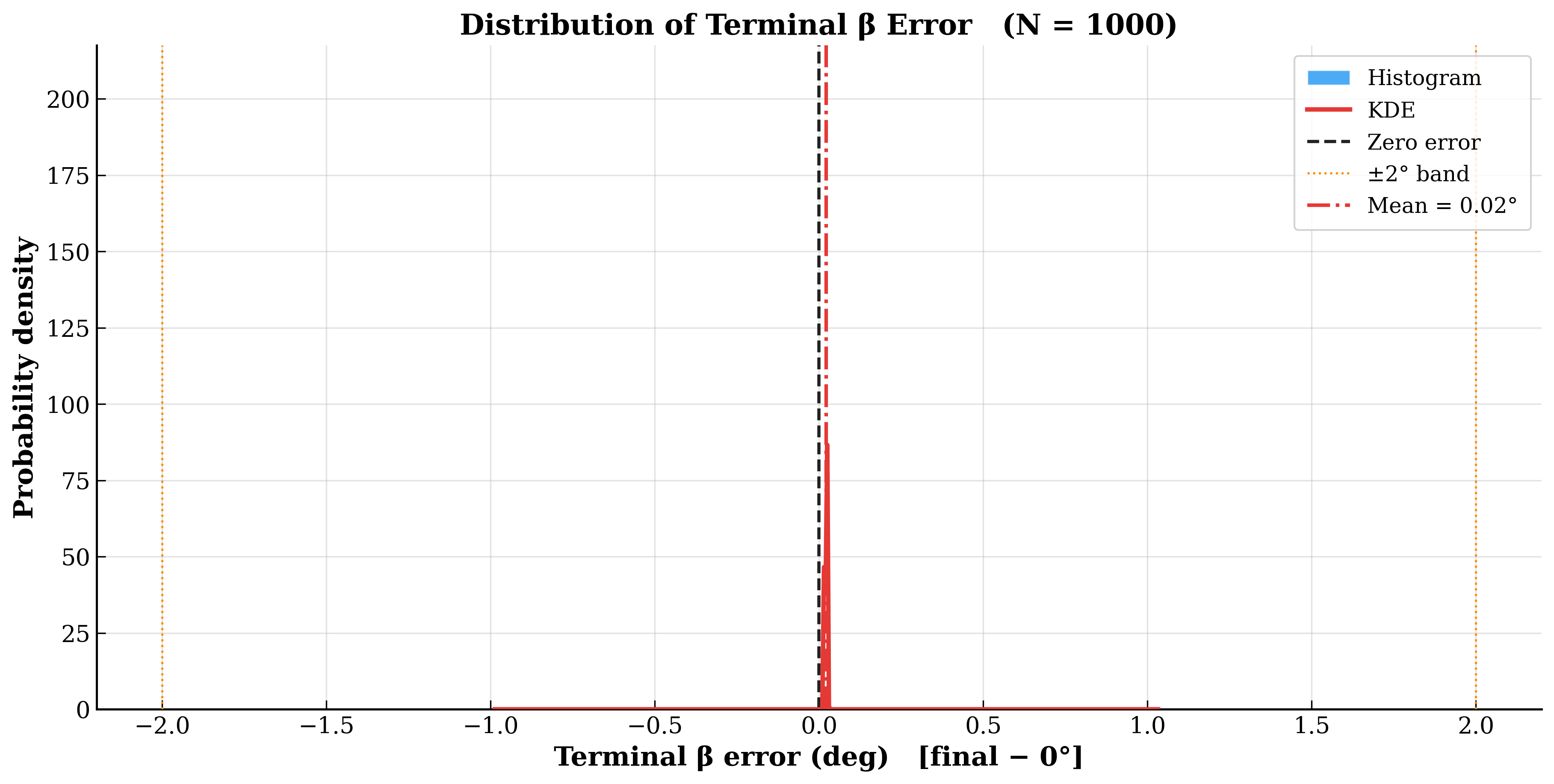}}
    \\
    \subfloat[Terminal $\mu$ error]{
        \includegraphics[width=0.98\linewidth]{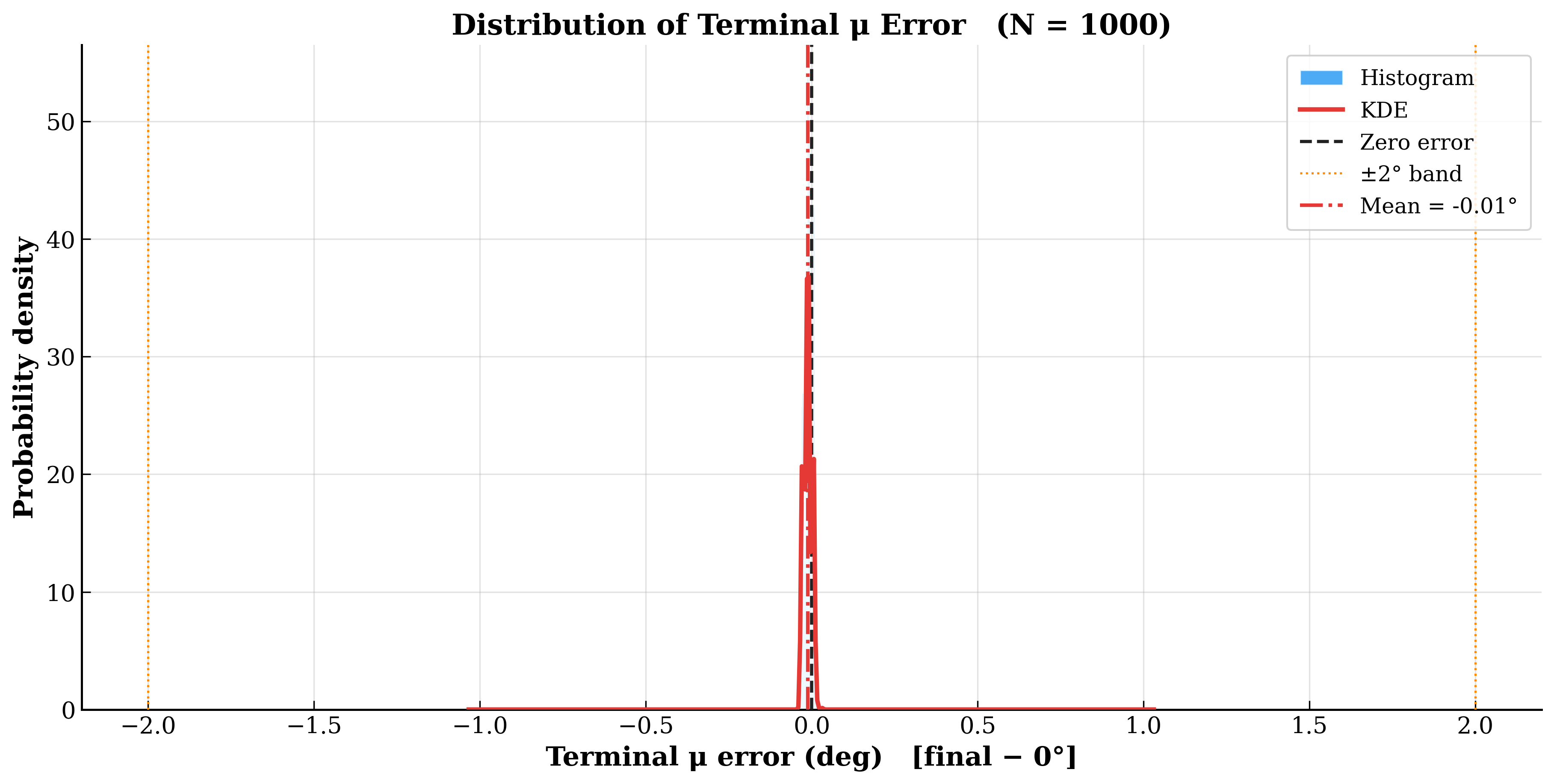}}
    \caption{Distribution of terminal tracking error across the $N=1000$ Monte Carlo trials for each controlled output.}
    \label{fig:mc_error_dist}
\end{figure}

This constitutes an empirical robustness check over the tested initial-condition envelope, rather than a formal sensitivity or reachability analysis over model uncertainty, aerodynamic parameters, or disturbance realizations. For readability, angles in this subsection are reported in degrees, consistent with the tolerance bands discussed below; elsewhere in the paper, angles follow the radian convention of Table~\ref{tab:f18_state_bounds_rad} and the simulation environment.

Fig.~\ref{fig:mc_trajectories} shows the resulting trajectory bands for the three controlled outputs. In all three channels, the 25--75th percentile band collapses onto the target trajectory well before $t=10$~s regardless of the randomized initial condition, and the 5--95th percentile band, which captures the most extreme sampled initial conditions, converges shortly after. The sideslip and bank-angle channels (Figs.~\ref{fig:mc_trajectories}b--c) exhibit larger transient excursions for initial conditions far from equilibrium, consistent with the cross-axis coupling discussed in Section~IV-A, but both settle to their zero target with no steady-state offset.

Fig.~\ref{fig:mc_error_dist} quantifies terminal tracking accuracy across all $N$ trials. The terminal angle-of-attack error (Fig.~\ref{fig:mc_error_dist}a) is concentrated within $\pm 0.5^\circ$ of the target, well inside the $\pm 2^\circ$ tolerance band, with a mean error of $-0.21^\circ$. The terminal sideslip and bank-angle errors (Figs.~\ref{fig:mc_error_dist}b--c) are effectively zero across the entire sample ($-0.01^\circ$ to $0.02^\circ$ mean), confirming that the controller consistently returns these channels to their equilibrium target irrespective of the randomized initial condition.

\begin{figure}[H]
    \centering
    \makebox[\linewidth][c]{%
        \includegraphics[width=1.1\linewidth]{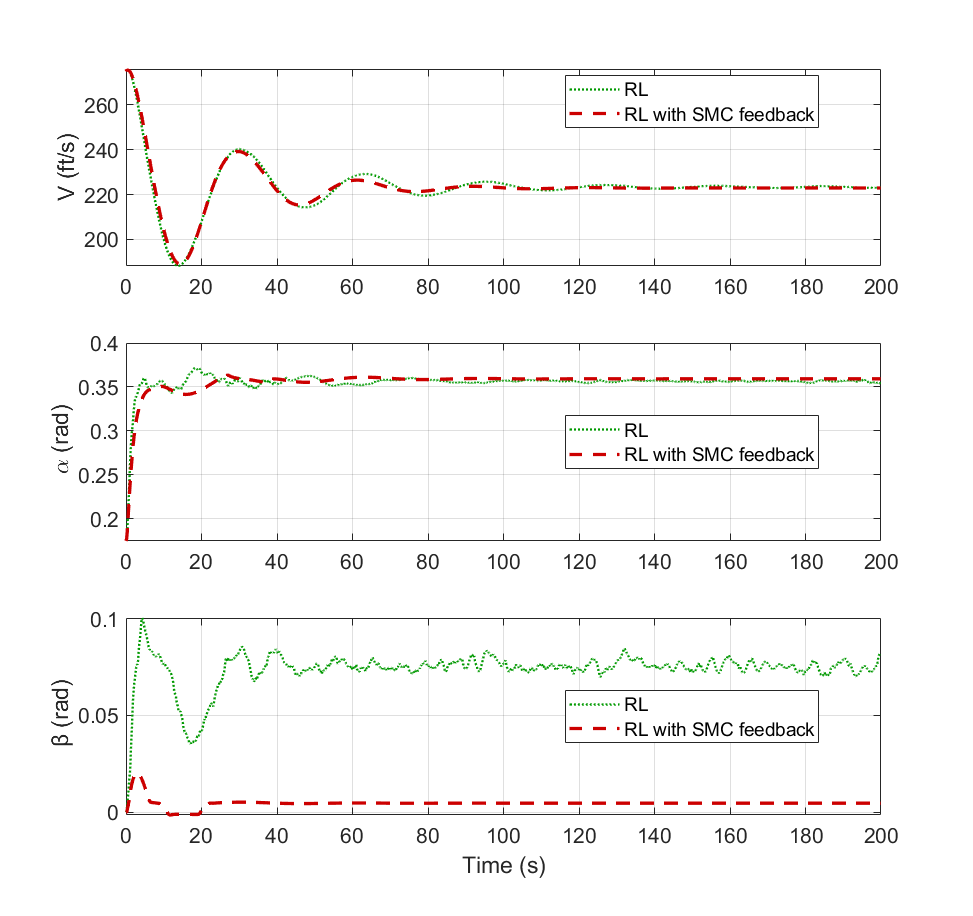}}
    \caption{System response for a partially trained PPO policy when target angle of attack is $0.349$ rad ($20^o$): \\
    (a) velocity ($V$) (ft/sec) (b) angle of attack ($\alpha$) (rad) (c) angle of sideslip ($\beta$) (rad).}
    \label{fig:20_vab}
\end{figure}

\subsection{Partially Trained RL Model: Effect of SMC Safety Layer}

A Proximal Policy Optimization (PPO) agent was trained in the F-18/HARV upset-recovery environment for 10 million timesteps. The policy is not trained to achieve single target from a fixed initial condition but it was trained to track a range of target angle of attack $\alpha$ in the range 0 to 0.698 radians ($0^{\circ}$ to $40^{\circ}$) from any random initial condition, therefore it retains noticeable variability and occasional destabilizing actions in unseen and unexplored states.

To assess the safety-layer effect:
\begin{itemize}
    \item \textbf{RL (partial trained PPO):} direct policy output applied to the plant, subject only to actuator saturation limits \cite{r14,r15}.
    \item \textbf{RL with SMC feedback:} the same partial policy filtered through the proposed SMC feedback loop, which overrides the RL output whenever stability is at risk.
\end{itemize}

\begin{figure}[h]
    \centering
    \makebox[\linewidth][c]{%
        \includegraphics[width=1.1\linewidth]{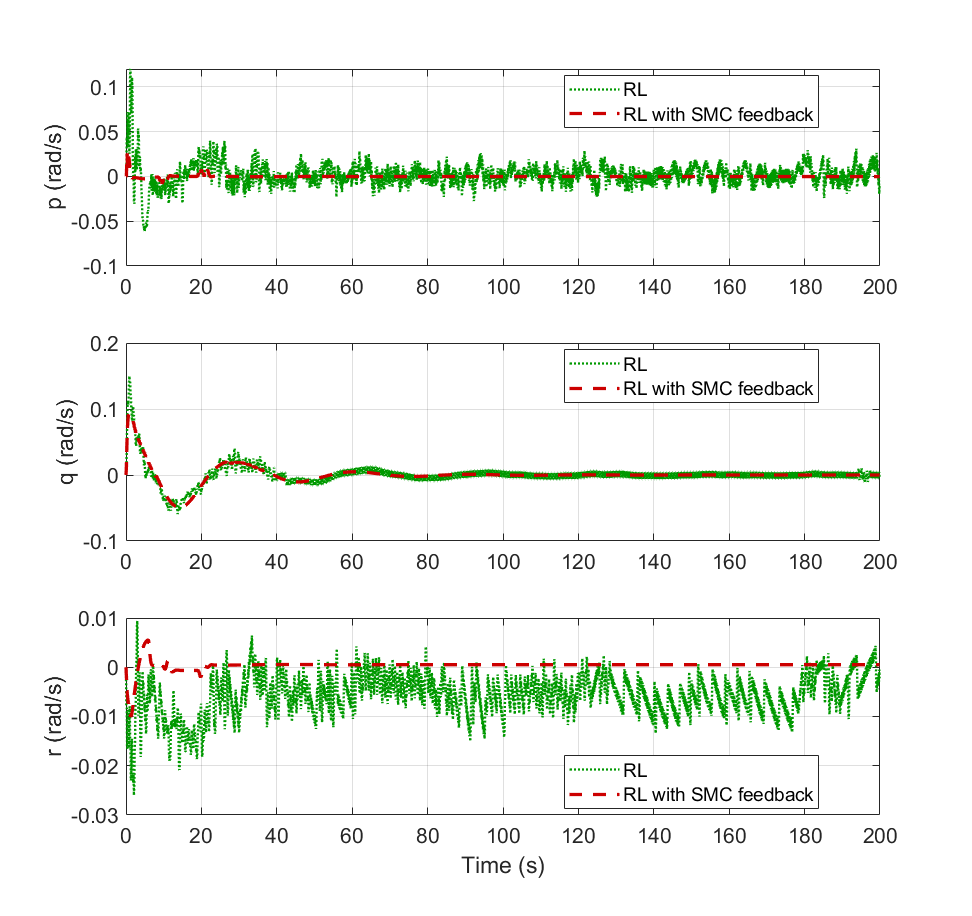}}
    \caption{System response for a partially trained PPO policy when target angle of attack of $0.349$ rad ($20^o$) \\
    (a) roll rate ($p$) (rad/sec) (b) pitch rate ($q$) (rad/sec) (c) yaw rate ($r$) (rad/sec).}
    \label{fig:20_pqr}
\end{figure}

\begin{figure}[H]
   \centering
    \makebox[\linewidth][c]{%
        \includegraphics[width=1.1\linewidth]{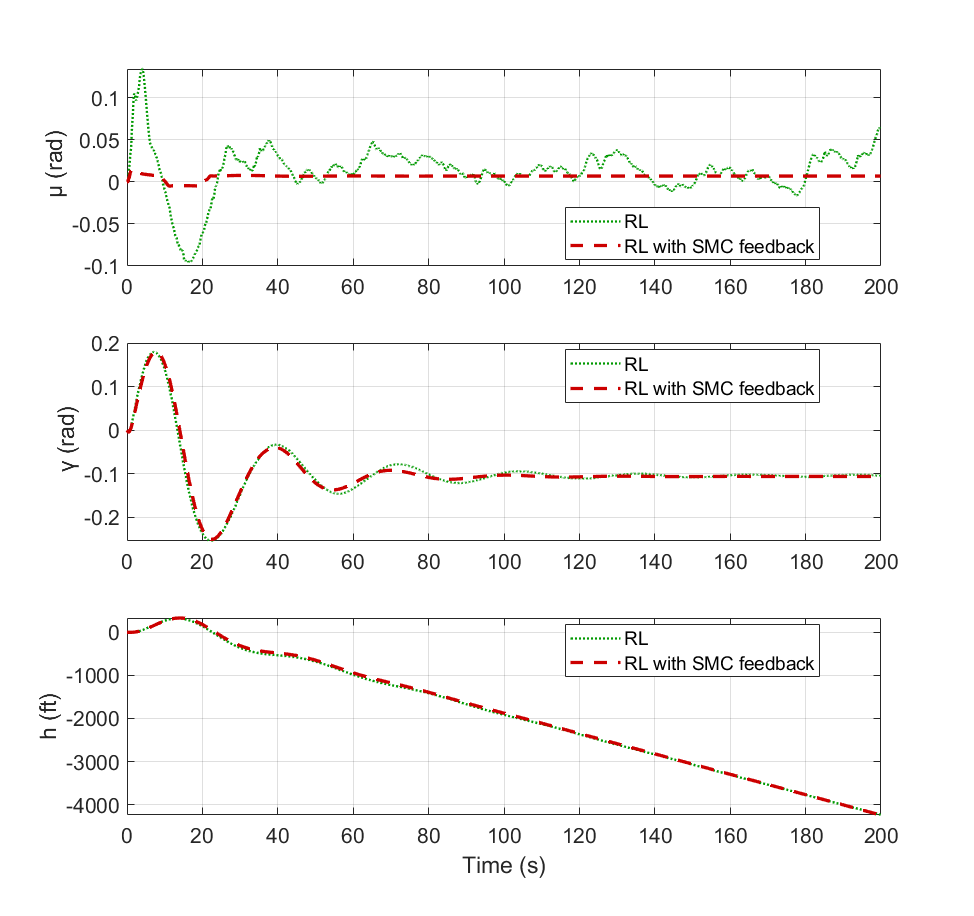}}
    \caption{System response for a partially trained PPO policy when target angle of attack of $0.349$ rad ($20^o$) \\
    (a) bank angle ($\beta$) (rad) (b) flight path angle ($\gamma$) (rad) (c) altitude ($h$) (ft).}
    \label{fig:20_mgh}
\end{figure}

\begin{figure}[H]
    \centering
    \makebox[\linewidth][c]{%
        \includegraphics[width=1.1\linewidth, height=0.4 \textheight]{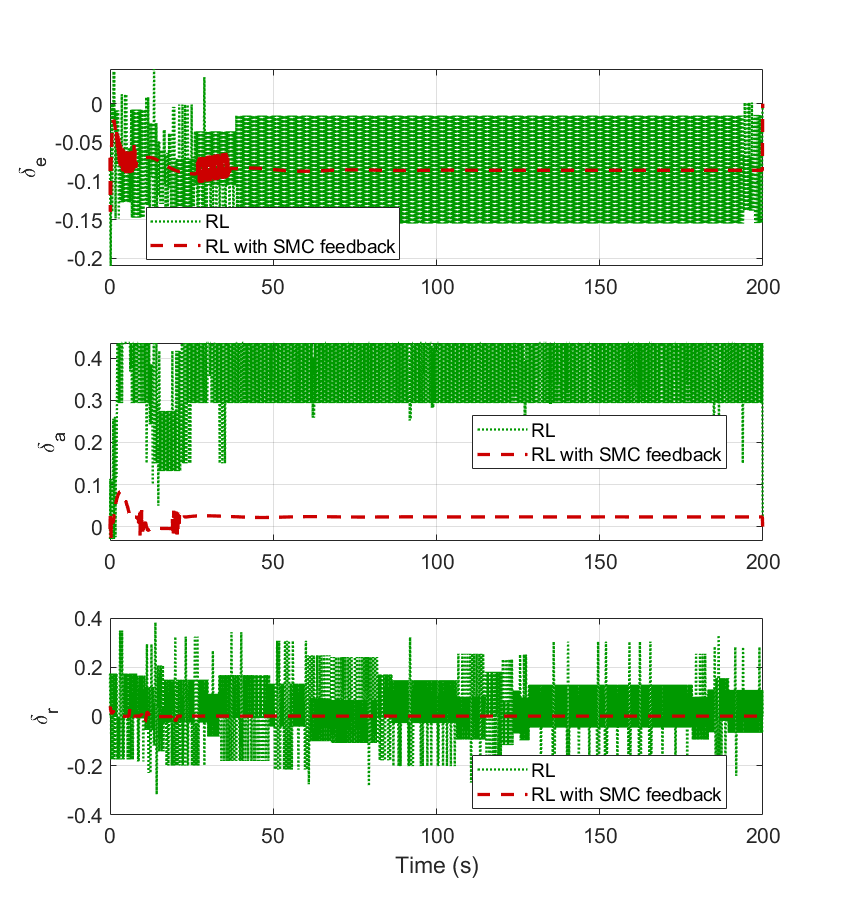}}
     \caption{Control surface deflection values generated by partially trained PPO policy when target angle of attack of $0.349$ rad ($20^o$) \\
    (a) elevator ($\delta_e$) (rad) (b) aileron ($\delta_a $) (rad) (c) rudder ($\delta_r $)(rad).}
    \label{fig:20_u}
\end{figure}

Figs. \ref{fig:20_vab} - \ref{fig:20_u} (time histories of states and control-surface deflections) show that in RL-only simulation, the aircraft exhibits oscillatory trajectories, with $\alpha$ overshoots in Fig. \ref{fig:20_vab}(b), small but significant offset error in angle of sideslip in Fig. \ref{fig:20_vab}(c), yaw rates persisting for a long time period in Fig. \ref{fig:20_pqr}(c). In RL simulation with SMC feedback case, the overshoot is comparatively  reduced, the rates settle to zero, and control usage remains smooth, despite the underlying policy being only partially converged. The control inputs in the RL-only case exhibit significant chattering, whereas the addition of SMC feedback substantially suppresses these high-frequency oscillations, making them practically negligible as seen in Fig. \ref{fig:20_u}.

The same partially trained model was also evaluated for different target angles of attack, $0.17$~rad ($10^\circ$) $0.52$~rad ($30^\circ$) etc to assess generalization beyond the training condition; the results were consistent with the $20^\circ$ case above, with the safety layer again suppressing overshoot and chattering relative to the RL-only policy.

It can be observed from this simulation that the safety layer suppresses destabilizing control deflections without noticeably impeding beneficial learned responses, provided the RL action lies within safe bounds. 

Even with incomplete training, the hybrid scheme can salvage and enhance policy performance by enforcing stability envelopes, making it a viable intermediate step in the deployment pipeline when full training is ongoing.

\subsection{Fully Trained RL Model Simulation: RL vs.\ SMC vs.\ RL with SMC feedback}
In the third phase, the RL policy was fully trained for spin recovery over $120$ million timesteps, with domain randomization across mass, aerodynamic coefficients, and wind disturbances. The training objective combined the two-phase reward described in Section~IV-A with potential-based reward shaping to promote rapid rate damping followed by attitude stabilization. The aircraft was allowed to remain in spin for first 30 seconds of simulation on purpose to demonstrate the spin condition. After this interval, controller was allowed to take up the recovery task. This is done to demonstrate the aircraft behavior in the spin condition for first 30 seconds. The network architecture is an Actor-Critic Network of [256, 128] hidden layers - Tanh configuration each. Learning rate and Discount factor were kept at 5e-5 and 0.99 to encourage smooth learning and long-term rewards. Following the tube-based switching logic for $\zeta$ described in Section~V, the control authority factor is set to $\zeta = 1$ when $\lvert e_\alpha\rvert$, $\lvert e_\beta\rvert$, or $\lvert e_\mu\rvert$ exceed $0.05$, while inside this bounded error tube $\zeta = 0$ and the sliding-mode feedback is relaxed, allowing the reinforcement-learning policy to dominate the control action within certified bounds.

\begin{figure}[H]
    \centering
    \includegraphics[width=1\linewidth]{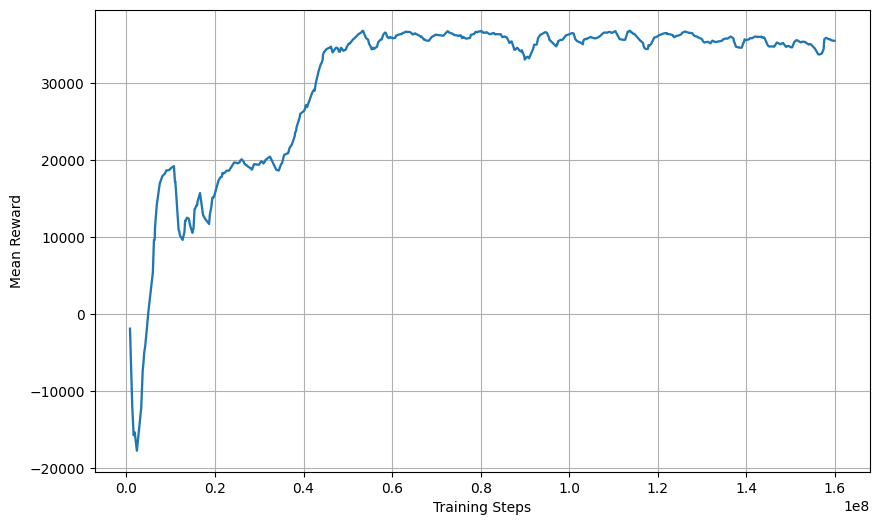}
    \caption{Mean episodic reward during PPO training of the fully trained
    spin-recovery policy.}
    \label{fig:training_reward}
\end{figure}

Fig.~\ref{fig:training_reward} shows the mean episodic reward recorded during
PPO training of this fully trained policy over the $120$ million ($1.2\times10^8$)
timesteps used for the reported results. Reward exhibits an initial
destabilizing dip during early exploration, before the Phase~1 rate-damping
term (Section~IV-A) comes to dominate the shaped reward, followed by rapid
improvement and convergence to a stable plateau of approximately
$33{,}000$--$37{,}000$ after roughly $5\times10^7$ steps. This indicates that
training had converged well before the timestep budget used to obtain the
policy reported in this section. All results in this section correspond to a
single training run with a fixed random seed; averaging results across
multiple independent training seeds was not performed and is left for future
work.

We do comparison of three controllers:
\begin{itemize}
    \item \textbf{RL-only} --- fully trained PPO policy, subject only to the actuator saturation limits of Table~\ref{tab:obs_action_space}, with no SMC feedback or safety filter.
    \item \textbf{SMC-only} --- pure robust feedback law without learned component.
    \item \textbf{RL control with SMC feedback} --- fully trained PPO policy with SMC feedback and the safety filter.
\end{itemize}

It should be noted that the RL-only baseline, as defined above, receives no
constraint enforcement beyond actuator saturation and does not benefit from
the SMC feedback loop or the safety filter present in the hybrid controller.
Consequently, differences observed between the RL-only and the RL--SMC hybrid
controller in this section (e.g., in control chattering or constraint
adherence) partly reflect the presence versus absence of this constraint
enforcement layer, rather than solely a difference in the underlying learned
policy. The comparison is nonetheless informative for the purpose of this
study, since it isolates the practical effect of adding the proposed SMC
safety layer on top of an otherwise identical trained policy, which is
precisely the deployment scenario the hybrid architecture targets; it is not,
however, intended as a controlled ablation that separates the contribution of
SMC feedback from that of constraint enforcement alone.

\begin{figure}[h]
    \centering
    \makebox[\linewidth][c]{%
        \includegraphics[width=1.2\linewidth]{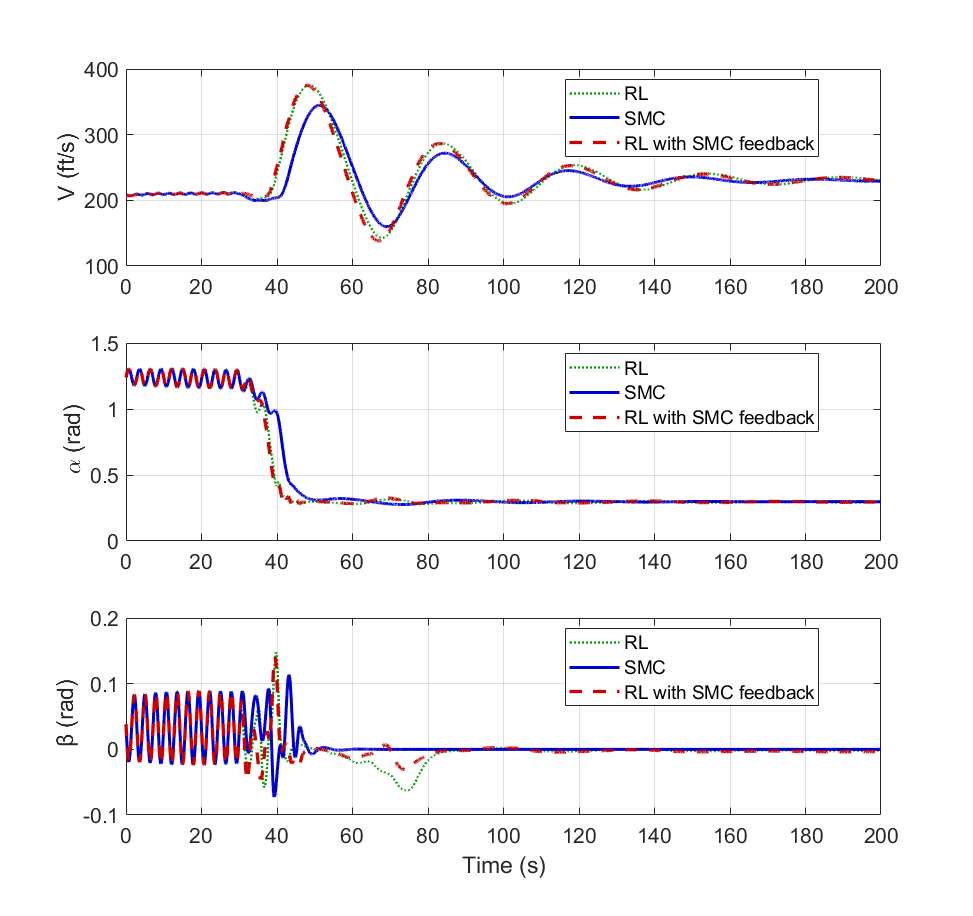}}
    \caption{System response of an aircraft under spin recovery using a trained PPO policy. \\
    (a) velocity ($V$) (ft/sec) (b) angle of attack ($\alpha$) (rad) (c) angle of sideslip ($\beta$) (rad).}
    \label{fig:comb_vab}
\end{figure}

\begin{figure}[h]
    \centering
    \makebox[\linewidth][c]{%
        \includegraphics[width=1.1\linewidth]{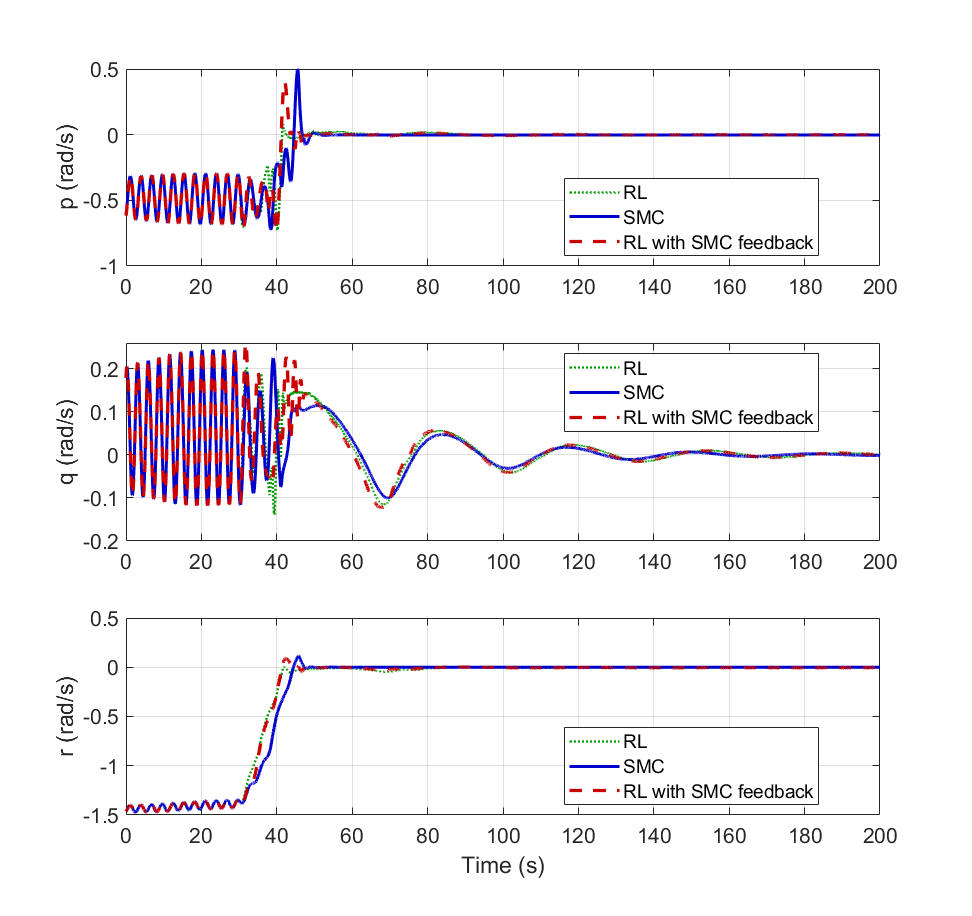}}
    \caption{System response for a trained PPO policy aircraft undergoing spin recovery \\
    (a) roll rate ($p$) (rad/sec) (b) pitch rate ($q$) (rad/sec) (c) yaw rate ($r$) (rad/sec).}
    \label{fig:comb_pqr}
\end{figure}

\begin{figure}[h]
    \centering
    \makebox[\linewidth][c]{%
        \includegraphics[width=1.1\linewidth]{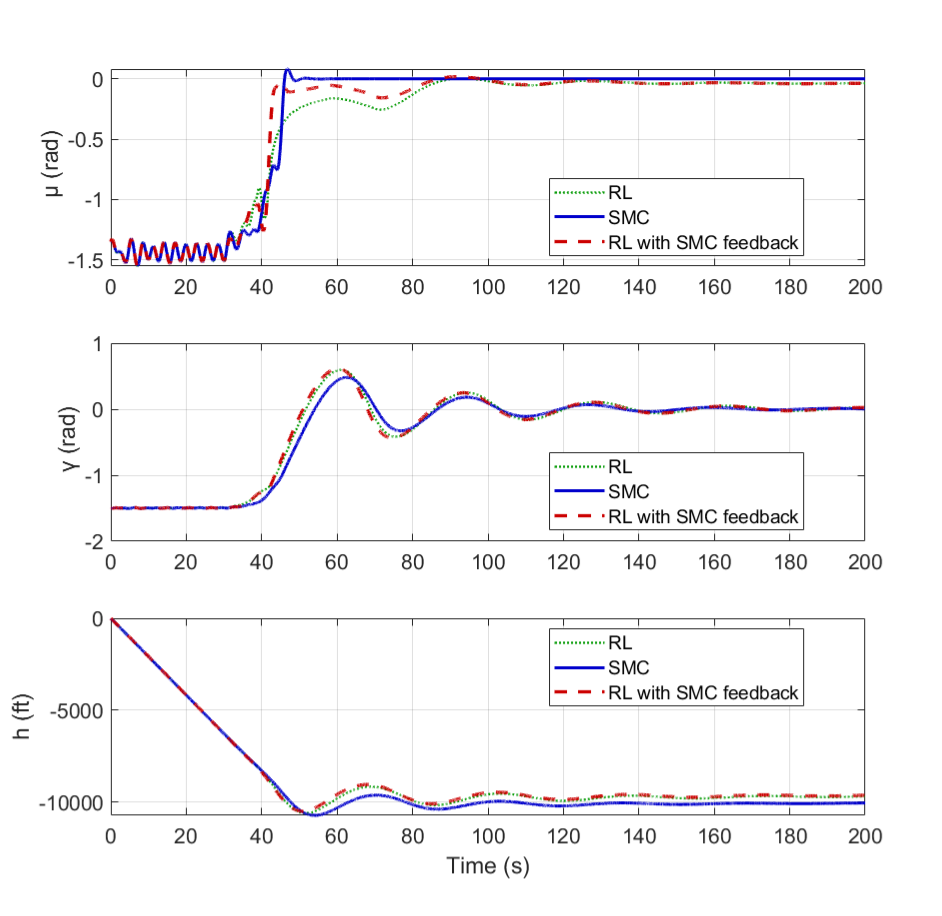}}
    \caption{System response for a trained PPO policy aircraft undergoing spin recovery \\
    (a) bank angle ($\beta$) (rad) (b) flight path angle ($\gamma$) (rad) (c) altitude ($h$) (ft).}
    \label{fig:comb_mgh}
\end{figure}

\begin{figure}[h]
    \centering
    \makebox[\linewidth][c]{%
        \includegraphics[width=1.12\linewidth]{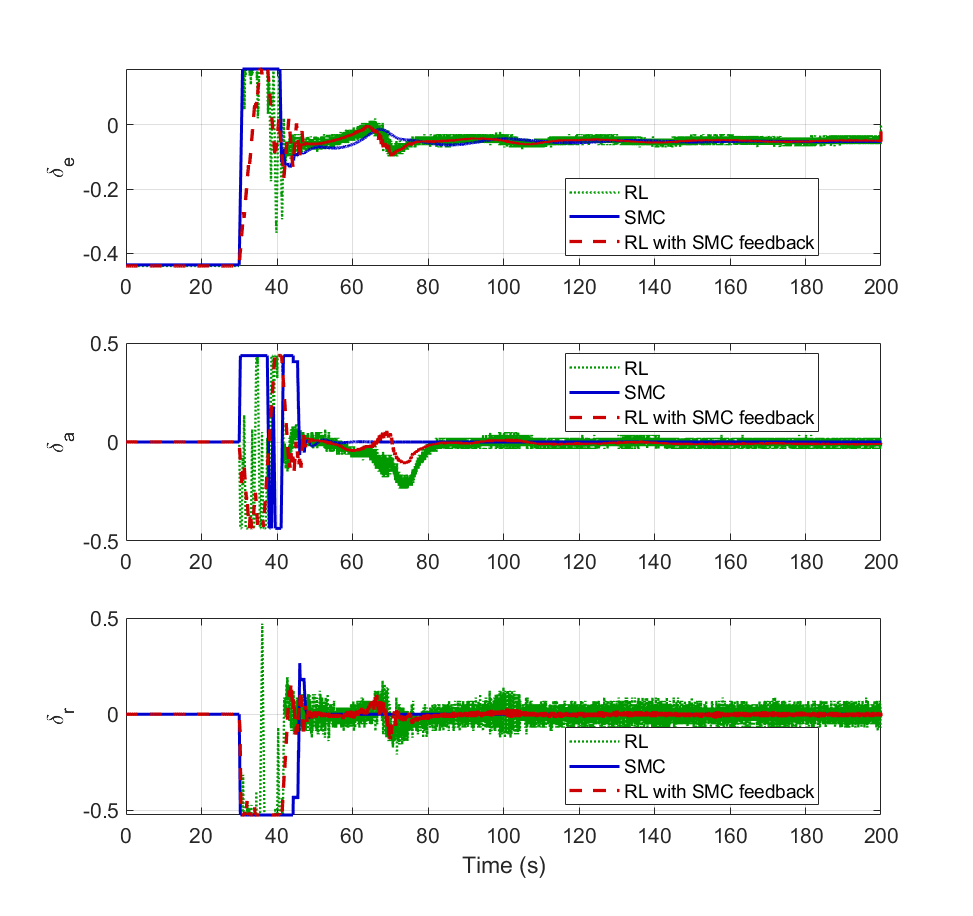}}
    \caption{Control surface deflection values generated by a trained PPO policy aircraft undergoing spin recovery \\
    (a) elevator ($\delta_e$) (rad) (b) aileron ($\delta_a $) (rad) (c) rudder ($\delta_r $)(rad).}
    \label{fig:comb_u}
\end{figure}

Figs. \ref{fig:comb_vab}--\ref{fig:comb_u} reveal the following trends:

 In Fig. \ref{fig:comb_vab}, The RL-only controller achieves the shortest recovery to the desired angle of attack $\alpha_d$, followed closely by the RL–SMC hybrid controller. The difference in recovery time is negligible, as the hybrid scheme exhibits similar convergence behavior with oscillations of negligible amplitude.
 
There is a significant delay of around 8-10 seconds for SMC as seen in Fig.\ref{fig:comb_vab}(b). From Fig. \ref{fig:comb_vab} RL with SMC produces least overshoots in angle of sideslip and even in velocity and flight path angle which is not directly being controlled.The RL controller achieves rapid trajectory tracking but displays mild
oscillatory behavior in $\mu$. The RL–SMC hybrid controller exhibits the small deviations in $\mu$; however, its overall behavior lies between that of the RL-only and SMC controllers.

 In the fully trained model, it can be observed that the RL controller gives the best performance when the model has been properly trained as shown in Fig. \ref{fig:comb_u}. However the control inputs produced are chattered and to train a RL model to produce controls with chattering can be very expensive computationally as it will require more training \cite{r28,r5}. It is also worth mentioning that even if such training is feasible it can pose the risk of overfitting phenomenon. SMC controller can be seen as an effective but a conservative controller which provides robustness and safety guarantees, something that RL model even though trained sufficiently may still lack \cite{doi:10.2514/1.50186}. RL control with SMC feedback filter produces inputs with significant reduction in chattering as well gives comparatively better performance that has faster tracking as well as safety guarantees because of its hybrid architecture.

\emph{Interpretation:} In the partially trained models, the hybrid architecture consistently outperforms individual controllers by leveraging the anticipatory, coupling-aware commands from RL while maintaining the constraint-respecting stability properties of SMC. The safety filter ensures that, even in off-nominal conditions, the RL term cannot destabilize the system, while the SMC term rapidly corrects any deviation. With fully trained RL model, the proposed hybrid architecture gives optimal performance as the safety layer regularizes commands, lowers total variation and chatter while retaining the coordinated character of the learned policy. One key aspect in which the RL controller outperforms the other approaches is its anticipatory nature, which motivates the use of RL as a learned feedforward control component. In contrast, the SMC controller is inherently reactive, and its corrective action is therefore slightly delayed when compared with the RL-based policy. The proposed hybrid architecture achieves both the anticipatory behavior of RL as well as stability guarantees of SMC. This necessitates careful tuning of the control authority of the SMC feedback term, denoted by $\zeta$.

\subsection{Key Insights}
The following key insights are observed from the results.
\begin{itemize}

    \item RL is an anticipatory controller so its response is much faster than SMC controller which is a reactive controller.
    \item Even partially trained RL policies benefit markedly from the SMC safety layer, which stabilizes residual oscillations and bounds overshoot without completely overriding useful learned behavior.
    \item The safety-filtered design ensures robustness to modelling errors and disturbances, enabling wider-envelope application without retraining.
    \item The ability to operate with partial RL policies under SMC safety constraints provides a practical road map for incremental deployment in safety-critical domains.
    \item Actuator usage with RL-only is aggressive and noisy; SMC and RL-SMC hybrid substantially reduce the actuator total variation, and high-frequency chatter.
    \item The hybrid control architecture preserves the anticipatory benefits of RL while enforcing SMC-style stability and constraint compliance, providing the best possible trade-off for flight-critical use.
\end{itemize}

\section{CONCLUSION}

This work develops and evaluates a learning-augmented flight-control architecture that combines a deep reinforcement learning policy as a learned nominal compensation component with a sliding-mode controller in feedback, guarded by a supervisory safety filter on surface deflections and rates. Unlike prior learning–augmented control frameworks, this work provides an explicit control-theoretic integration of reinforcement learning and sliding-mode control through a bounded learned feedforward compensation formulation. The learned policy is modeled as a matched, norm-bounded input, and a tunable control authority factor is introduced to regulate the dominance of the robust feedback law. A Lyapunov-based analysis establishes a direct relationship between admissible learning error, disturbance magnitude, and sliding-mode gain selection, yielding a certified invariant neighborhood under actuator saturation.

Across upset-recovery trials on the F-18/HARV model, the hybrid controller consistently delivered the best overall trade-off . It preserved the anticipatory behavior of the learned policy while enforcing stability and hard actuator constraints, thereby reducing overshoot, residual oscillations, and control chatter relative to either RL-only or SMC-only baselines. Notably, even partially trained policies became reliably usable when wrapped with the SMC safety layer, and fully trained policies benefited from regularization that lowered total variation in the commands without erasing coordinated responses.

Methodologically, the paper contributes a simple integration recipe that is implementation-ready: (i) a two-phase, PBRS-based reward that first damps rates and then tightens attitude tracking, (ii) an SMC design that supplies robustness to modeling errors and disturbances, and (iii) a lightweight safety filter that caps learned authority and enforces deflection and rate limits. A Lyapunov-style analysis clarifies how the feedback layer counteracts residual mismatch from the learned nominal compensation policy.

There are limitations as the results are simulation-based, actuator dynamics are simplified, and tuning choices for gains and safety limits were empirical. From an implementation standpoint, the proposed hybrid controller is lightweight at deployment time: online evaluation requires only a single forward pass through the trained actor network (two hidden layers of 256 and 128 units) to obtain $u_{\mathrm{RL}}$, combined with the closed-form SMC law of~\eqref{eq:total_input}, the tube-based hysteresis switch for $\zeta$, and the elementwise saturation filter, all of which involve only algebraic operations with no online optimization or iterative solves. This is in contrast to optimization-based alternatives such as MPC, which require solving a constrained optimization problem at every control step; the proposed architecture instead confines all optimization to offline training, making the online computational cost essentially fixed and small, and favorable for real-time implementation on embedded flight-control hardware, though this has not yet been verified on such hardware. In terms of tunability, the controller exposes a small number of interpretable parameters, the SMC gains $K$ and $\epsilon$, the sliding surface coefficients $\lambda_1,\lambda_2$, and the tube thresholds $e_{\mathrm{on}},e_{\mathrm{off}}$, each of which has a direct role in the stability condition~\eqref{eq:stability_condition} or the switching logic, rather than requiring re-tuning of the trained policy itself. Future work will migrate the architecture to hardware-in-the-loop, incorporate explicit actuator and sensor dynamics, quantify online computational cost and latency on representative embedded hardware, and investigate adaptive or reference-governor style scheduling of SMC gains and safety limits. Overall, the study shows that combining learned compensation with robust sliding-mode feedback and explicit constraint handling is a practical path to bring modern learning into safety-critical flight control.

\bibliographystyle{IEEEtran}
\bibliography{ref}

\begin{IEEEbiography}[{\includegraphics[width=1in,height=1.25in,clip,keepaspectratio]{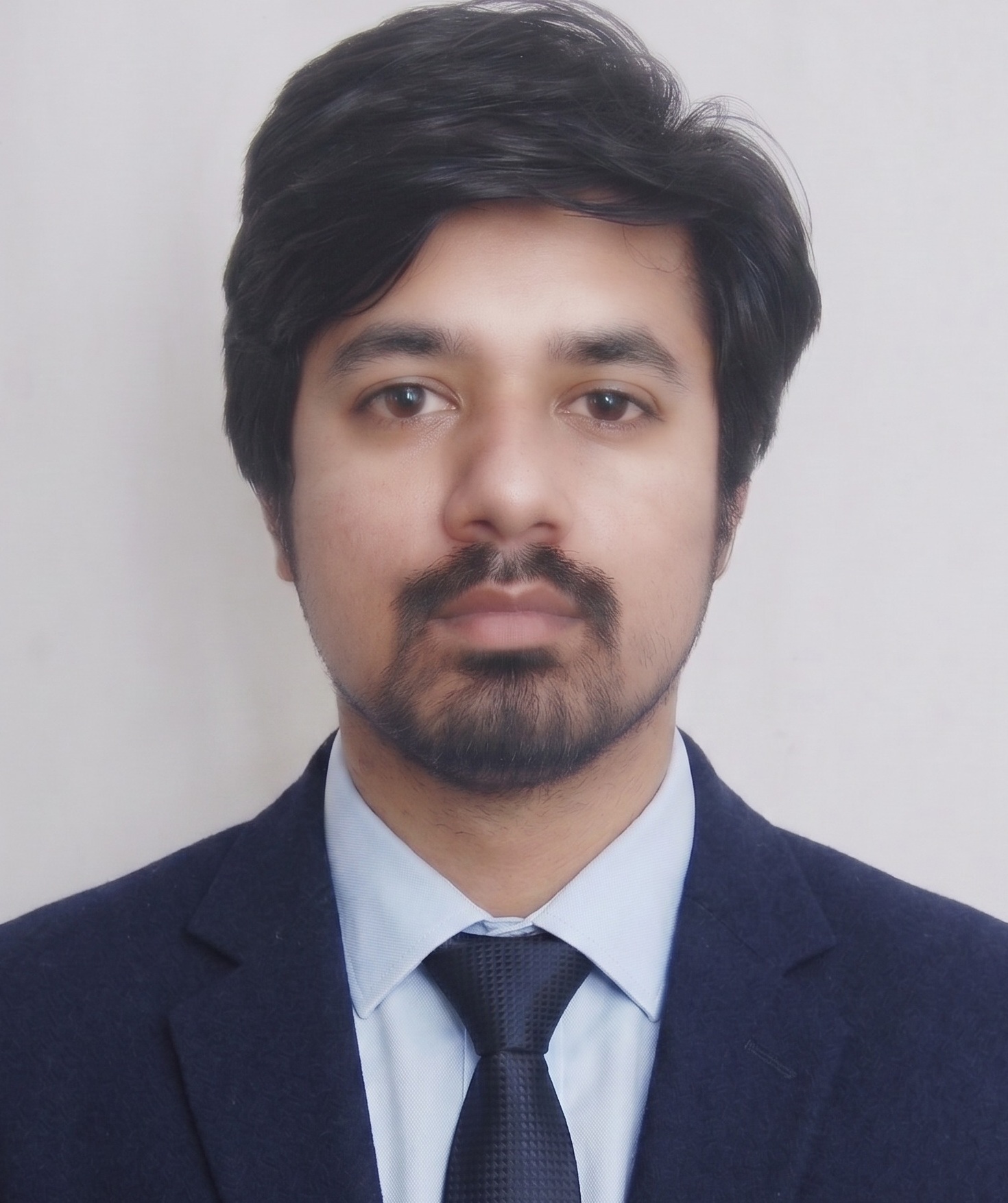}}]{Imran Sayyed}
received the Bachelor's degree in Mechanical Engineering from Panjab University and the M.S. degree in Aerospace Engineering from the Indian Institute of Technology Madras, Chennai, India.

His research focuses on nonlinear and adaptive control of aircraft, with emphasis on sliding mode control, reinforcement learning-based flight control, and six-degree-of-freedom flight dynamics modeling. His work involves developing robust controllers for spin recovery and integrating model-based and data-driven approaches for flight envelope protection.
\end{IEEEbiography}

\begin{IEEEbiography}[{\includegraphics[width=1in,height=1.25in,clip,keepaspectratio]{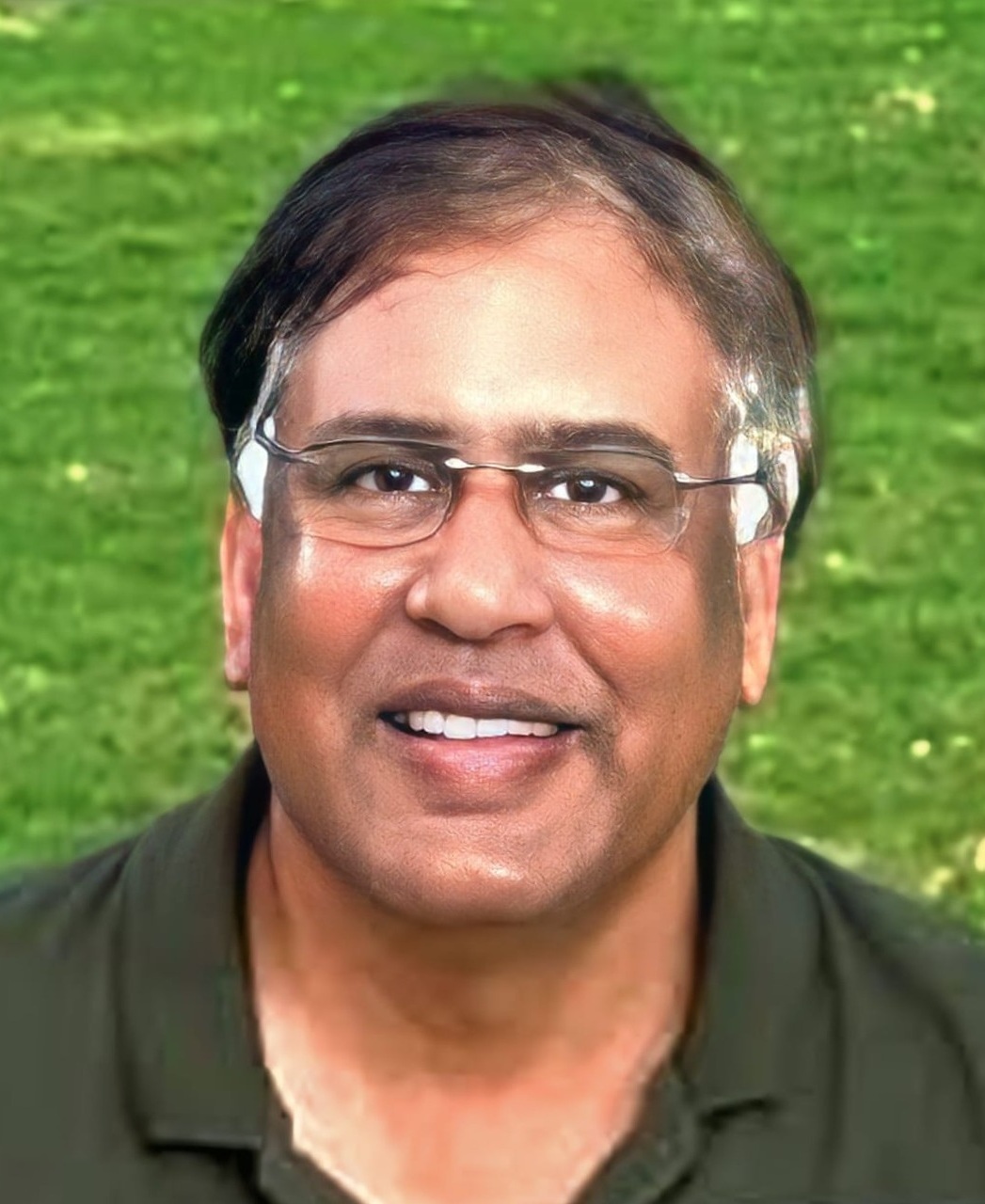}}]{Dr. Nandan Kumar Sinha}
is a Professor with the Department of Aerospace Engineering, Indian Institute of Technology Madras, Chennai, India. He received the B.Tech., M.Tech., and Ph.D. degrees in Aerospace Engineering from the Indian Institutes of Technology Bombay and Kanpur. Before joining IIT Madras as a faculty member in 2006, he was a Visiting Postdoctoral Scholar with Technische Universität Darmstadt, Germany.

His research interests include the dynamics and control of aerospace vehicles, flight mechanics, guidance and control, nonlinear control, and estimation. He has coauthored two books and has published extensively in these areas.
\end{IEEEbiography}

\end{document}